\documentclass[a4paper,twocolumn,11pt,accepted=2025-03-21]{quantumarticle}
\pdfoutput=1
\usepackage[utf8]{inputenc}
\usepackage[english]{babel}
\usepackage[T1]{fontenc}
\usepackage{amsmath}
\usepackage{hyperref}
\usepackage{amsfonts}
\usepackage{tikz}
\usepackage{multirow}
\usepackage[caption=false]{subfig}
\usepackage{bbm}
\usepackage{asymptote}
\usepackage{diagbox}
\usepackage{caption}
\usepackage{ragged2e}
\usepackage{manfnt}
\usepackage{lipsum}
\usepackage{wasysym}
\usepackage[numbers,sort&compress]{natbib}
\usepackage{epstopdf}
\usepackage{overpic}
\usepackage{braket}
\usepackage{algorithm}
\floatname{algorithm}{Protocol}
\usepackage{algpseudocode}
\usepackage{siunitx}

\newcommand{\bd}{\begin{displaymath}}

\newcommand{\nn}{\nonumber \\}

\newcommand{\Mref}[1]{M\ref{#1}}

\begin{document}

\title{Template demonstrating the quantumarticle document class}

\title{Fault-tolerant Quantum Error Correction Using a Linear Array of Emitters}

\author{Jintae Kim}
\affiliation{Department of Physics, Sungkyunkwan University, Suwon 16419, Korea}
\affiliation{Institute of Basic Science, Sungkyunkwan University, Suwon 16419, Korea}
\orcid{0000-0002-2762-2809}
\author{Jung Hoon Han}
\affiliation{Department of Physics, Sungkyunkwan University, Suwon 16419, Korea}
\author{Isaac H. Kim}
\email{ikekim@ucdavis.edu}
\affiliation{Department of Computer Science, University of California, Davis, CA 95616, USA}
\maketitle

\begin{abstract}
We propose a fault-tolerant quantum error correction architecture consisting of a linear array of emitters and delay lines. In our scheme, a resource state for fault-tolerant quantum computation is generated by letting the emitters interact with a stream of photons and their neighboring emitters. Depending on the number of emitters $n_e$, we study the effect of delay line errors in two regimes: when $n_e$ is a small constant of order unity and when $n_e$ scales with the code distance. Between these two regimes, the logical error rate steadily decreases as $n_e$ increases, from a scaling of $\exp(-c\eta^{-1/2})$ to $\exp(-c'\eta^{-1})$, where $\eta$ is the error rate per unit length in the delay line, for some constants $c,c'>0$. We also carry out a detailed study of the break-even point and the fault-tolerance overhead. These studies suggest that the multi-emitter architecture, using the state-of-the-art delay lines, can be used to demonstrate error suppression, assuming other sources of errors are sufficiently small.

\end{abstract}

\section{Introduction}
\label{sec:1}

One of the fundamental discoveries in quantum computation is the theory of fault-tolerant quantum computation~\cite{gottesman98, gottesman98a, gottesman09, kitaev03, aharonov08, knill98}. While realistic quantum computers are noisy, if their noise rate is below the fault-tolerance threshold, one can simulate the behavior of a noiseless quantum computer arbitrarily well using a noisy quantum computer. 

Recently, rapid progress in quantum computing technology led to an explosion of works on fault-tolerant quantum computing architectures~\cite{barends14, google19,harty14, ballance16, kim13, madsen22, lukin19, kaufman18, thompson19, endres19}. These studies aim to develop protocols tailored to specific hardwares, with the goal of maximizing the efficiency of the underlying error correction protocols. One of the most well-studied architecture is the surface-code based architecture, which are well-suited for planar array of superconducting qubits~\cite{fowler12}. Recent advances led to novel architectures suitable for implementing quantum low-density parity check codes~\cite{Tremblay2022,Delfosse2021,Strikis2023,Xu2023}. These theoretical advances were also accompanied by recent experimental milestones that demonstrate quantum error correction~\cite{Acharya2022,RyanAnderson2021,RyanAnderson2022,Bluvstein2023}. 

However, near-term quantum computers are still not powerful enough to carry out commercially useful quantum computations. A commonly cited commercial application of quantum computation is the study of challenging molecules such as FeMoCo~\cite{Reiher2016}. However, in spite of the progress made in quantum algorithms, the number of $T$-gates --- the most expensive fault-tolerant gate due to the costly magic state distillation --- is still estimated to be between $10^9$ and $10^{10}$, with the number of logical qubits estimated to be at least $10^3$~\cite{Lee2020,vonBurg2020}. 

Manufacturing and controlling such large number of qubits can pose a significant challenge. However, it is possible to mitigate this challenge by exploiting the unique physics provided by certain platforms. One such approach is to use a single-photon emitter such as quantum dot~\cite{Lodahl2015,senellart2017high,warburton2013single,michler2017quantum}. With photons, a promising approach is to build a cluster state, which is a resource state for measurement-based quantum computation~\cite{raussendorf01,raussendorf03} as an avenue for doing quantum computation quite distinct from those based on unitary circuits. Preparation of cluster state using photons is a well-studied subject. Many protocols have been studied both theoretically~\cite{nielsen04, rudolph10, lukin17, wan21, shi21, Bombin2021Interleaving, Bartolucci2021, Tzitrin2021, paesani23, Bourassa2020, lindner09} and experimentally~\cite{walther05,yokoyama13,furusawa16,lindner16,andersen19,furusawa19,ferreira22,ra23}. 

For instance, it is well-known that an emitter can be used to create a one-dimensional cluster state~\cite{lindner09}. By using a single delay line,  a single emitter can prepare a two-dimensional cluster state~\cite{lukin17,ferreira22}, which can be used for universal measurement-based quantum computation~\cite{raussendorf03}. 

With an additional delay line, one can even prepare a three-dimensional cluster state~\cite{wan21}, which is a resource state for universal \emph{fault-tolerant} quantum computation~\cite{harrington05,goyal06,goyal07,ra23}. In this scheme, all that is required for building a fault-tolerant logical qubit is a single emitter, two delay lines, and a single-photon detector. Compared to the alternatives which would require controlling hundreds if not thousands of physical qubits, the demand on the number of experimental components is more modest.

However, there is an important caveat. The error rate one needs to demonstrate error suppression puts a challenging demand on the quality of the delay line. Assuming that the error rate per unit length in the delay line is $\eta$, it was shown in Ref.~\cite{wan21} that the logical error rate scales as $\exp(-\eta^{-\frac{1}{2}})$, provided that the other error rates (e.g., two-qubit gates between the emitter and the photon) smaller than a threshold. Unfortunately, assuming the circuit-level noise is a depolarizing noise of strength of $0.1\%$, the existing state-of-the-art (loss) error rates reported in the commercial delay lines~\cite{Tamura2018} were below the break-even point~\cite{wan21}. More precisely, if one were to use such a delay line, logical error rate after error correction is expected to be strictly higher than the circuit-level error. As such, while the approach of Ref.~\cite{wan21} was conceptually appealing, it did not seem practical considering the current experimental capabilities.

\begin{figure}[t]
\centering
\includegraphics[width=0.48\textwidth]{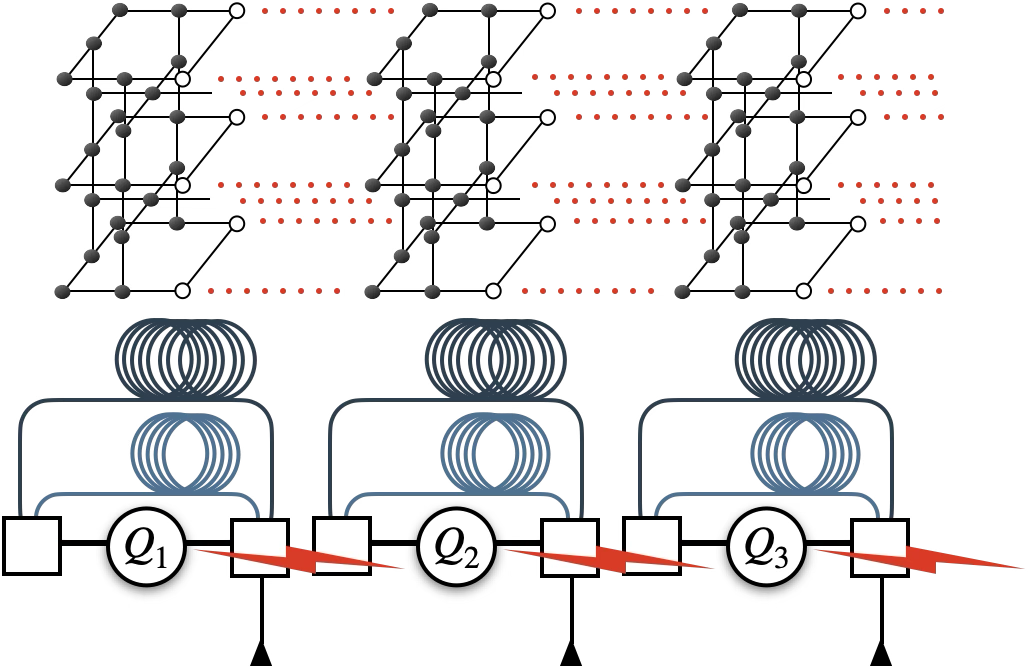}
\caption{Schematic description of our scheme, which is a linear array of emitters and delay lines. Each emitter $Q_i$ is connected to two delay lines, designated as navy and blue lines, through which the photons propagate. The detectors and the interactions between neighboring emitters are depicted using triangles and thunder signs, respectively. Each emitter is responsible for building a thin ``slab'' of the cluster state located directly above it. (The cluster state depicted in the figure is a three-dimensional cluster state; however, other cluster states can be constructed in general.) The black dots represent photons within the slab, while the white dots indicate the positions of the corresponding photons in a different slab. Entanglement between the slabs of the cluster state, represented by red dotted lines, can be generated by applying entangling gates to adjacent emitters.}
\label{fig:setup}
\end{figure}

The main purpose of this paper is to study a generalization of the scheme in Ref.~\cite{wan21} that can mitigate this issue. Our scheme involves emitters that are arranged on a linear array. What we envision is a scheme similar to the one proposed in Ref.~\cite{wan21}, but including extra emitters that can interact with their neighboring emitters. A schematic description of our approach is shown in Figure~\ref{fig:setup}. We note that such a gate can be applied between a pair of gate-defined semiconductor quantum dots~\cite{Xue2021,Noiri2022,Mills2022,Madzik2022,Philips2022}, and therefore it may also be possible to apply it between other types of semiconductor quantum dots that are quantum emitters ~\cite{senellart2017high,warburton2013single,michler2017quantum}.


Generally speaking, as the number of emitters increase, the performance of our scheme (quantified in terms of the lowest logical error rate one can achieve) improves. We discuss these improvements on three fronts. First, if the number of emitters $n_e$ is a constant of order unity (e.g., 2 or 3), the logical error rate scales as $\exp(-\sqrt{n_e/\eta})$. While this may seem like a modest improvement over the $\exp(-\eta^{-1/2})$ scaling in Ref.~\cite{wan21}, we emphasize that the logical error rate scales \emph{exponentially} with these numbers. Therefore, increasing $n_e$ by a factor of few can lead to a factor of few \emph{orders of magnitude} change in the logical error rate. We observe this phenomena in our numerical studies.

Second, if the number of emitters is large, we obtain a more favorable scaling form for the logical error rate. If $n_e$ is linear in the code distance $d$, we find the logical error rate scales as $\exp(-m^{-1}\eta^{-1})$ (instead of $\exp(-\eta^{-\frac{1}{2}})$), where $m=d/n_e$. Between this regime and the regime in which $n_e = O(1)$, the logical error rate steadily improves as $n_e$ increases. 

Our improvement stems from the fact that the time each photon experiences in the delay line can be reduced as we increase the number of emitters. Thus by increasing the number of emitters, one effectively reduces the amount of error each photon is experiencing in the delay line. This explains why, as we increase the number of emitters, the logical error rate improves. This suggests that increasing the number of emitters is generally favorable, provided that the two-qubit error rates between the neighboring emitters are sufficiently low.



Physically, the entangling gate between emitters may rely on a mechanism different from the ones used for gates between emitters and photons. One possible approach involves using superconducting qubits interacting with microwave photons or phonons~\cite{mirhosseini2019cavity,ferreira2020collapserevivalartificialatom,ferreira22}. Alternatively, one may use quantum dots, using the long-range gates between quantum dots mediated by cavity~\cite{QDotQED} or waveguide~\cite{tiranov2023collective}\footnote{While there are alternative ways of realizing a gate between quantum dots that demonstrated high fidelity~\cite{Volz2014,Sipahigil2016,Goban2014,Tiecke2014,Reiserer2014}, to the best of our knowledge, it is not clear if the same quantum dot can be used as emitters.}. Experimentally, two-qubit gate fidelities exceeding 99\% have been demonstrated in systems with a small number of gate-defined quantum dots ~\cite{Xue2021,Noiri2022,Mills2022,Madzik2022}. However, scaling such systems to a larger number of quantum dots often results in reduced fidelity\footnote{However, see Ref.~\cite{Philips2022} for the recent advance in scaling the system to up to six quantum dots.}. As a result, a potential concern for our scheme lies in the practical challenge of scaling the system while maintaining high two-qubit gate fidelities.

However, we show in this paper that the two-qubit error rate one can tolerate is significantly higher than what one might naively expect. We consider an anisotropic error model in which the ratio between the two-qubit error rate between the emitters and the other gates can be varied. When $n_e=O(1)$, we find that the thresholds of our schemes remain practically unchanged even if we assume that the two-qubit error rate between the emitters is almost ten times larger than the other error rates. When $n_e$ is large, the thresholds of our schemes do depend on the error ratio, though still in the range of $2.5\times 10^{-3}\sim 4.0 \times 10^{-3}$, similar to the thresholds of the single-emitter protocols~\cite{wan21}. Thus our scheme is tolerant against such experimentally motivated noise models.

The rest of the paper is structured as follows. In Section \ref{sec:2}, we provide an executive summary of a cluster state preparation protocol which involves a single emitter and two delay lines. This is similar to the one proposed in Ref. \cite{wan21}, but modified in a way that is suitable for the generalization presented in Section~\ref{sec:3}. In Section \ref{sec:3}, we propose several cluster state preparation protocols using multiple emitters and delay lines. In Section \ref{sec:4a}, we study the threshold of our scheme by varying the number of emitters over several different error models, in the absence of delay line error. We find the threshold to remain largely intact, independent of the number of emitters, even if the gates between the emitters are far noisier than the other gates. In Section \ref{sec:5}, we extend the analysis in Section~\ref{sec:4a} to the setup in which the delay line error is present. In Section \ref{sec:optimal_delay_line}, we addressed the improvement of a multi-emitter protocol in comparison to a single-emitter protocol. In Section~\ref{sec:6}, we conclude with a discussion. In particular, we focus on the prospect of achieving error suppression using realistic experimental parameters, using the multi-emitter architecture we propose.

\section{Single emitter protocols}
\label{sec:2}

In this Section, we provide an executive summary of a protocol for preparing a three-dimensional (3D) cluster state using a single emitter. While the main body of our work follows that of Ref.~\cite{wan21}, we also make some changes that are suitable for our generalization to the multi-emitter protocols in Section~\ref{sec:3}.

Physically, the emitter can be an atom or a quantum dot~\cite{Lodahl2015,senellart2017high,michler2017quantum}. However, our exposition will simply view them as a qubit, interacting with photons, which can be viewed as another qubit. To differentiate the two, we shall refer to the emitter as the \emph{ancilla qubit} and the photons as \emph{data qubits}. From this perspective, the operations we apply are well-known gates such as controlled-$Z$ ($\mathsf{CZ}$), controlled-NOT ($\mathsf{CNOT}$), and the Hadamard acting on the ancilla qubit. These gates can be only applied between the ancilla qubit and a data qubit, not between data qubits. (We remark that our $\mathsf{CNOT}$ gate is slightly different from the standard one in that it acts as a $\mathsf{CNOT}$ gate only for a specific subset of quantum states; see Eq.~\eqref{eq:id}~\cite{wan21} and the surrounding texts for more details.)

\begin{figure}[ht]
\centering
\includegraphics[width=0.55\columnwidth]{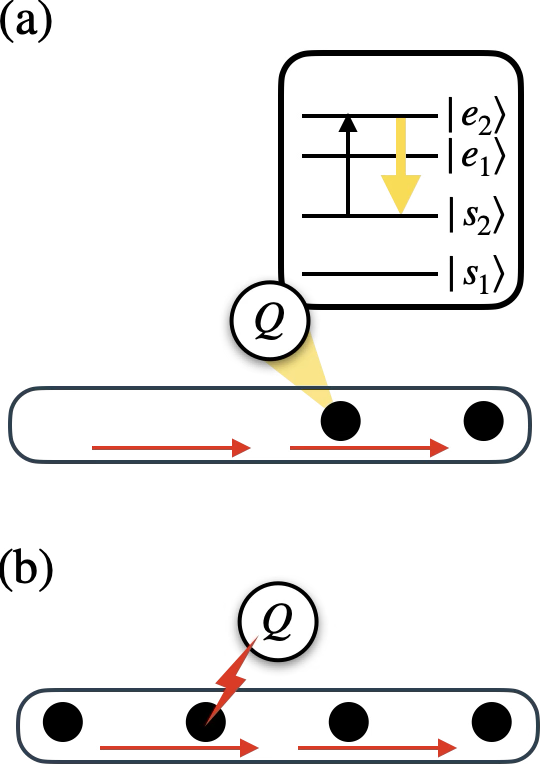}
\caption{A schematic description of an emitter ($Q$) and a delay line is presented. The data qubits (black dots) propagate through the delay line. (a) The emitter has both stable ($\ket{s_1},~ \ket{s_2}$) and radiative states ($\ket{e_1},~ \ket{e_2}$). A data qubit can be emitted into the delay line (yellow arrow) following a transition from a stable state to a radiative state (black arrow). (b) The emitter can interact with the data qubit through scattering.}
\label{fig:emitter}
\end{figure}

Experimentally, these gates are implemented in the following way. (For the following discussion, note that the emitter is a multi-level system with both stable and radiative states.) First, the initialization of the data qubits correspond to creation of photons. This is based on a rapid resonant excitation pulse induces a transition from a stable state to a radiative state, leading to an emission of a photon into the delay line [Figure~\ref{fig:emitter}(a)]. Second, the two-qubit gate between the data qubit and the ancilla qubit is based on the scattering of the photon off the emitter. This process can induce certain two-qubit gates, such as $\mathsf{CZ}$ and $\mathsf{CNOT}$ ~\cite{Volz2014, Sipahigil2016, Goban2014, Tiecke2014}; see [Figure~\ref{fig:emitter}(b)] for a schematic description. Lastly, the Hadamard gate on the ancilla qubit can be realized by applying a pulse to the emitter.

The main purpose of using these gates is to ultimately prepare some \emph{cluster states}, also known as graph states. This state is defined in terms of a graph $G=(V,E)$. Here $V$ is a set of vertices and $E$ is a set of (undirected) edges. We will represent the edge between vertices $i,j\in V$ as $\{ i,j\} = \{ j, i\}$. Then, the cluster state $|\psi_G\rangle$ is defined as 
\begin{align}
|\psi_G \rangle=\prod_{(i,j) \in E} Z_{i,j} \bigotimes_{i' \in V} |+\rangle_{i'},\label{eq:psiG}
\end{align}
where $\ket{+}_{i'}$ is the +1 eigenstate of the Pauli-$X$ operator acting on vertex $i'$, and $Z_{i,j}$ is the $\mathsf{CZ}$ gate on the $\{ i , j\}$ edge, satisfying $Z_{i,j}\ket{s_i s_j}=(-1)^{s_is_j}\ket{s_i s_j}$. The state $\ket{s_i}$ satisfies $Z_i \ket{s_i}=(-1)^{s_i} \ket{s_i}$, where $s_i=0,1$ and $Z_i$ is a Pauli-$Z$ operator acting on the site $i$. Note that the vertices in the graph only involve the data qubits. The ancilla qubit only assists in creating the target cluster state $|\psi_G\rangle$; it is not part of the qubits that constitute $|\psi_G\rangle$. The cluster state can be described also in terms of the stabilizers~\cite{hein06, griffiths08}. The canonical generating set of the stabilizer group is  $\left\{X_i\prod_{j:\{i,j\}\in E} Z_j: \{i,j\} \in E\right\}$,  where we denoted Pauli-$X$ and Pauli-$Z$ on vertex $i$ as $X_i$ and $Z_i$, respectively.

\subsection{Cluster state construction from a single emitter}
\label{sec:2A}


Here we explain a general approach to prepare a cluster state using a single emitter, following the discussion in Ref.~\cite{wan21}. We define a sequence of subgraphs $G[k] \subset G$ and a quasi-subgraph $G[k]'$ of the graph $G$ as follows:
\begin{align}
G[k] &\equiv ([k] ,E[k]),\nn
G[k]' &\equiv ([k] \cup \{Q\},E[k] \cup \{\{Q,k\}\}),
\label{eq:G-and-G'} 
\end{align}
where $[k] = \{1,\cdots, k\}$ is a subset of data qubits whose elements are labeled by a non-negative integer, and $E[k] = \{\{i,j\} \in E:i,j \in [k]\}$ is the set of edges among the data qubits in $[k]$. When representing the graph $G$ as an adjacency matrix, the subgraph $G[k]$ corresponds to the $k \times k$ submatrix of the full adjacency matrix. The quasi-subgraph $G[k]'$ contains, in addition to the vertices and edges in $G[k]$, a vertex corresponding to the ancilla qubit $Q$ and an additional edge between $Q$ and the $k$-th qubit. 

The high level picture of our protocol is following: At each step, the quantum operation transforms $G[k-1]'$ into $G[k]'$. Specifically, the procedure involves entangling the previously introduced data qubits, $1$ through $k-1$, with the ancilla qubit if they should be entangled with $k$, wherein the ancilla qubit assumes the role of the next data qubit to be introduced. Once the entanglement is established, the ancilla qubit is transferred to the data qubit $k$ using a $\mathsf{SWAP}$ gate. The simple example is the following~\cite{wan21}:
\begin{align}
&\text{SWAP}_{Q,2} Z_{Q,1}|+\rangle_{Q} \otimes |+\rangle_{2} \otimes |+\rangle_{1} \nn
& = Z_{1,2}|+\rangle_{Q} \otimes |+\rangle_{2} \otimes |+\rangle_{1}, \label{eq:CZ}
\end{align}
where $Q$ is the ancilla qubit and ${\rm SWAP}_{a,b}$ is the $\mathsf{SWAP}$ gate acting on qubits $a$ and $b$. Subsequently, we will replace the $\mathsf{SWAP}$ gate with the aforementioned set of gates.

Therefore, in general, the cluster state obtained at the $k$-th step is related to the one at the $(k-1)$-th step as follows~\cite{wan21}
\begin{align}
 \ket{\psi_{G[k]'}} = & Z_{Q,k}\text{SWAP}_{Q,k} \Bigl[\prod_{i:\{i,k\} \in E[k]} Z_{Q,i}\Bigl]\nn
&\times Z_{Q,k-1} \ket{\psi_{G[k-1]'}} \otimes \ket{+}_k.\label{eq:step}
\end{align}

\begin{figure}[ht]
\centering
\includegraphics[width=0.9\columnwidth]{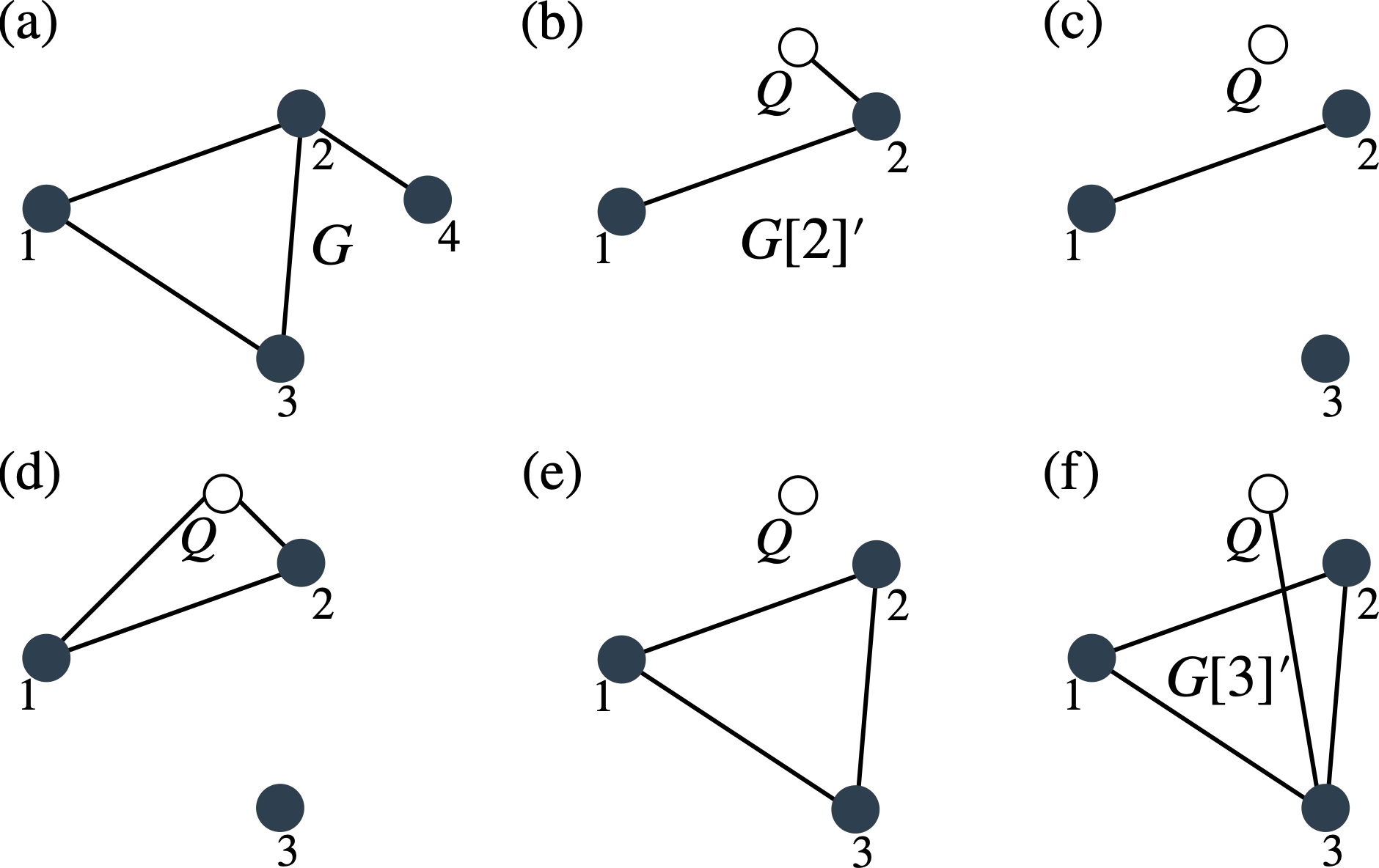}
\caption{(a) The graph $G$ is illustrated. (b)-(f) A procedure for generating the state corresponding to the graph $G[3]'$ from the state corresponding to the graph $G[2]'$. }
    \label{fig:G}
\end{figure}

For the example of graph $G$ in Figure~\ref{fig:G}(a), the construction process from $G[2]'$ [Figure~\ref{fig:G}(b)] to $G[3]'$ [Figure~\ref{fig:G}(f)] is as follows: applying $\mathsf{CZ}$ gate between $Q$ and $2$ [Figure~\ref{fig:G}(c)], applying $\mathsf{CZ}$ gates between $Q$ and $2$ and between $Q$ and $1$ [Figure~\ref{fig:G}(d)], applying $\mathsf{SWAP}$ gate between $Q$ and $3$ [Figure~\ref{fig:G}(e)], and finally applying $\mathsf{CZ}$ gate between $Q$ and $3$ [Figure~\ref{fig:G}(f)].

Now, let us replace the $\mathsf{SWAP}$ gate with the aforementioned set of gates. The standard approach to implement the $\mathsf{SWAP}$ gate is to convert it into three $\mathsf{CNOT}$ gates. However, if the dominant source of error are the two-qubit gates (which is often the case), it is desirable reduce the number of $\mathsf{CNOT}$ (or equivalently, $\mathsf{CZ}$ gate). In such circumstances, it would be more advantageous to use the following decomposition, which only uses one $\mathsf{CNOT}$ gate~\cite{wan21}:
%
\begin{align}
Z_{Q,j} \text{SWAP}_{Q,j} \ket{\phi}_Q \otimes \ket{+}_j=H_Q X_{Q,j} \ket{\phi}_Q \otimes \ket{0}_j,\label{eq:id}
\end{align}
where $X_{Q,j}$ is a $\mathsf{CNOT}$ gate with the ancilla as the control qubit and $\ket{\phi}_Q$ is an arbitrary state of $Q$. In our setup, the states on which the $\mathsf{CNOT}$ gate acts always take the form $|\phi\rangle_Q \otimes |0\rangle_j$. Therefore, in practice, the effective $\mathsf{CNOT}$ gate which satisfies Eq.~\eqref{eq:id} would be enough to construct the graph~\cite{wan21}. Thus we can sequentially build up the cluster state, arriving at $|\psi_{G[|V|]'}\rangle$. (Here $|V|$ is the number of vertices.)

To that end, we introduce a unitary $U_{k}$:
\begin{align}
U_{k}=H_Q X_{Q,k} \Bigl[ \prod_{i:\{i,k\} \in E[k]} Z_{Q,i}\Bigl] Z_{Q,k-1}, \label{eq:step2}
\end{align}
and rewrite $\ket{\psi_{G[k]'}}$ in Eq. \eqref{eq:step} as
\begin{align}
\ket{\psi_{G[k]'}}&= U_{k} \ket{\psi_{G[k-1]'}} \otimes \ket{0}_k\nn
&=\Bigl[\prod_{j=1}^k U_{j} \Bigl]\ket{+}_Q \bigotimes_{i=1}^k \ket{0}_i  . \label{eq:U1}
\end{align}
The state $\ket{\psi_{G[k]'}}$ is obtained by applying $U_1$ through $U_k$ on the initial product state $\ket{+}_Q \bigotimes_{i=1}^k \ket{0}_i$. At the very end, we apply $Z_{Q, |V|}$ to disentangle the ancilla from the data qubit and obtain $\ket{\psi_{G}}$:
\begin{align}
Z_{Q,|V|}\ket{\psi_{G[|V|]'}}=\ket{+}_Q\otimes\ket{\psi_{G}} . \label{eq:U2}
\end{align}
This completes the generation of the desired cluster state $|\psi_G\rangle$.

\begin{algorithm}[H]
\renewcommand{\thealgorithm}{S1}
\caption{Cluster state construction with one emitter}
\label{alg1}
\small
\begin{algorithmic}[1]
\State initialize $Q$ in $\ket{+}$
\For{$k=1$ to $|V|$}
    \State initialize qubit $k$ in $\ket{0}$
    \If{$\{k-1,k\} \in E$} \Comment{}
        \State apply $H_Q X_{Q,k} \prod_{i\neq k-1 \& \{i,k\} \in E[k]} Z_{Q,i}$ \Comment{}
    \Else \Comment{}
        \State apply $H_Q X_{Q,k} Z_{Q,k-1} \prod_{i:\{i,k\} \in E[k]} Z_{Q,i}$\Comment{}
    \EndIf \Comment{}
\EndFor
\State apply $Z_{Q,|V|}$
\end{algorithmic}
\end{algorithm}
\begin{algorithm}[H]
\renewcommand{\thealgorithm}{S2}
\caption{Cluster state construction with one emitter and its measurement and reinitialization}
\label{alg2}
\small
\begin{algorithmic}[1]
\State initialize $Q$ in $\ket{+}$
\For{$k=1$ to $|V|$}
    \State initialize qubit $k$ in $\ket{0}$
    \State apply $H_Q X_{Q,k} \prod_{i<k-1 \& \{i,k\} \in E[k]} Z_{Q,i}$ \Comment{}
    \If{$\{ k,k+1\} \notin E$ or $k=|V|$}\Comment{}
        \State apply $Z_{Q,k}$\Comment{}
        \State measure $Q$ in the Z-basis\Comment{}
        \State re-initialize $Q$ in $\ket{+}$\Comment{}
    \EndIf\Comment{}
\EndFor
\end{algorithmic}
\end{algorithm}

\subsection{Protocols}
\label{sec:protocolss}

There were several protocols proposed in Ref.~\cite{wan21} for cluster state preparation. We found that some of those are readily generalized to the multi-emitter setup whereas some are not. We introduce two protocols --- Protocol~\ref{alg1} and Protocol~\ref{alg2}--- that are easily adaptable to the multi-emitter case.\footnote{For the readers familiar with Ref. \cite{wan21}, we remark that our Protocol~\ref{alg1} is in fact Algorithm 2 of Ref.~\cite{wan21}. Also, Protocol~\ref{alg2} is a slight variant of Protocol B in Ref.~\cite{wan21}; here we apply $Z_{Q,k}$ prior to measuring $Q$ in the Z-basis whereas in Ref.~\cite{wan21}, they apply $Z_{k}$ after the measurement, specifically when the measurement outcome is $\ket{1}$.}  (Here the Roman letter S stands for the \emph{single} emitter.) The main difference between the two protocols is that Protocol~\ref{alg1} is purely unitary whereas Protocol~\ref{alg2} involves measurements and re-initialization of $Q$. 

We remark that the conditional statement in the protocols are included to minimize redundant operations, thereby avoiding extra unnecessary errors. More precisely, if  $\{k-1,k\} \in E$ belongs to the graph $E$, $U_k$ in Eq. (\ref{eq:step2}) involves the application of $Z_{Q,k-1}$ twice. This is because the qubit $k-1$ is linked to the ancilla qubit $Q$ in the quasi-subgraph $\ket{\psi_{G[k-1]'}}$. (See Figure~\ref{fig:G}, which illustrates the case where two $Z_{Q,2}$ operations have been applied.) Consequently, executing the optimized version of $U_k$ (lines 4-8 in Protocol~\ref{alg1}) results in fewer errors. In Protocol~\ref{alg2}, lines 4-9 correspond to the optimized $U_k$ for that protocol.

At first, one might worry that these protocols involve interaction of the ancilla qubit with an extensive number of data qubits. This can potentially lead to a single error on the ancilla qubit propagating to an extensive number of qubits. However, a remarkable fact first discovered in Ref.~\cite{wan21} is that their protocols do not lead to such adverse propagation of errors. A similar analysis can be carried out for our protocols. Also in our protocols one can find that the circuit-level one- or two-qubit errors propagate to, up to stabilizers, errors of constant weight.

\subsection{Fault-tolerant error correction}
\label{sec:2c}

\begin{figure}[ht]
\centering
\includegraphics[width=0.32\textwidth]{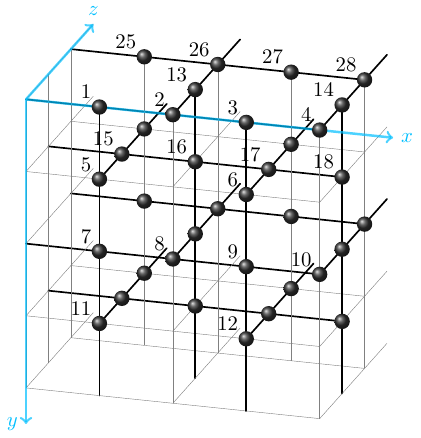}
\caption{A part of the $4 \times 4 \times 4$ 3D cluster state under periodic boundary condition is depicted. The numbering of the data qubits reflects the way the 3D cluster state is constructed by successively entangling additional data qubits.}
\label{fig:2}
\end{figure}

The $L\times L \times L$ 3D cluster state under periodic boundary condition for $L=4$ and the numbering of data qubits are depicted in Figure~\ref{fig:2}. The numbering is designed to construct the 3D cluster state by sequentially forming $xy$-planes. During the construction of each $xy$-plane, the rows along the $x$-axis are generated in succession. This numbering ensures that the maximum absolute value of the difference between the numbers of two connected qubits is approximately $L^2$, which is important for maintaining uniform and low delay line errors for each qubit.

An interesting direction for future work is to explore the structure of the cluster state and the optimal numbering of data qubits that can enhance efficiency and fault tolerance in the emitter system. Given that the geometry of fault-tolerant cluster states is closely tied to the delay line error for each qubit, investigating structures where the delay line error scales as less than $O(L^2)$ presents a valuable avenue for further research.

In this work, we mainly focus on the 3D cluster state, but other graph states and their numbering schemes could also be considered. For foliated quantum codes~\cite{stace16}, which are foliations of CSS codes, a similar numbering strategy can be used as described earlier, where the construction of each CSS code must be completed before starting the next. This would similarly ensure that the maximum absolute value of the difference between the numbers of two connected qubits is approximately equal to the number of qubits in a CSS code.

Once a (noisy) 3D cluster state is prepared, one can measure all the qubits in the $X$-basis. These measurement outcomes can be then fed into a decoder, which returns a correction. By studying whether the error and the correction form a logical operator or not, one can decide whether a logical error occurred or not~\cite{preskill02,harrington05,goyal06,goyal07}. We carried out such a numerical simulation using Protocols~\ref{alg1} and~\ref{alg2}, the result of which is presented below. (We assumed periodic boundary condition for both Protocols~\ref{alg1} and~\ref{alg2}.) Throughout this paper we used the minimum-weight perfect matching (MWPM) decoder, employing the open source code PyMatching~\cite{pymatching} and Stim~\cite{stim}. 

We note that performing numerical calculations for the 3D cluster state under periodic boundary conditions in all three directions is less practical in realistic scenarios, as the cluster state is intended to represent a quantum channel with input and output states localized on opposite temporal boundaries~\cite{harrington05}. However, to simplify numerical simulations and primarily investigate the bulk properties of the 3D cluster state, we adopt this boundary condition.

\subsubsection{Circuit-level noise}
\label{sec:circuit_level_noise}
We first study the performance of Protocol \ref{alg1} and \ref{alg2} under the standard circuit-level depolarizing noise. This is the standard model in which a single- and two-qubit depolarizing noise is applied after a single- and two-qubit gate is applied:
\begin{equation}
\begin{aligned}
{\cal D}_a^{(p)} (\rho)&=(1-p)\rho+\frac{p}{3}\sum_{P \in \{X,Y,Z\}} P_a \rho P_a \\
{\cal D}_{a,b}^{(p)} (\rho)&=(1-p)\rho+\frac{p}{15}\sum_{\substack{P,P' \in \{I,X,Y,Z\} \\ (P,P')\neq (I,I)}} P_b' P_a  \rho P_a P_b',
\end{aligned}\label{eq:dep}
\end{equation}
where $I_a$, $X_a$, $Y_a$, and $Z_a$ are the identity and the three Pauli gates acting on the vertex $a$. Here $\mathcal{D}_a^{(p)}$ is the single-qubit depolarizing channel of strength $p$ applied to qubit $a$ and $\mathcal{D}_{a,b}^{(p)}$ is the two-qubit depolarizing channel over qubit $a$ and $b$. For simplicity, we assume that the parameter $p$ remains the same for both single-qubit and two-qubit gates.

Against this noise model, we estimated the logical error rate $\overline{p}$ for a 3D cluster state of size $L \times L \times L$, where $L$ is even to satisfy periodic boundary condition. We averaged over $10^5$ realizations for each choice of $p$ and $L$. We estimate the logical error rate by fitting the data to a quadratic scaling ansatz~\cite{wan21}
\begin{align}
\overline{p}=\alpha+\beta(p-\mu)d^{1/\nu}+\gamma(p-\mu)^2d^{2/\nu},\label{eq:ansatz}
\end{align}
where $d=L/2$ and $(\alpha, \beta , \mu , \nu)$ are determined by fitting the data.

\begin{figure}[ht]
\centering
\small
\begin{overpic}[width=0.6\columnwidth]{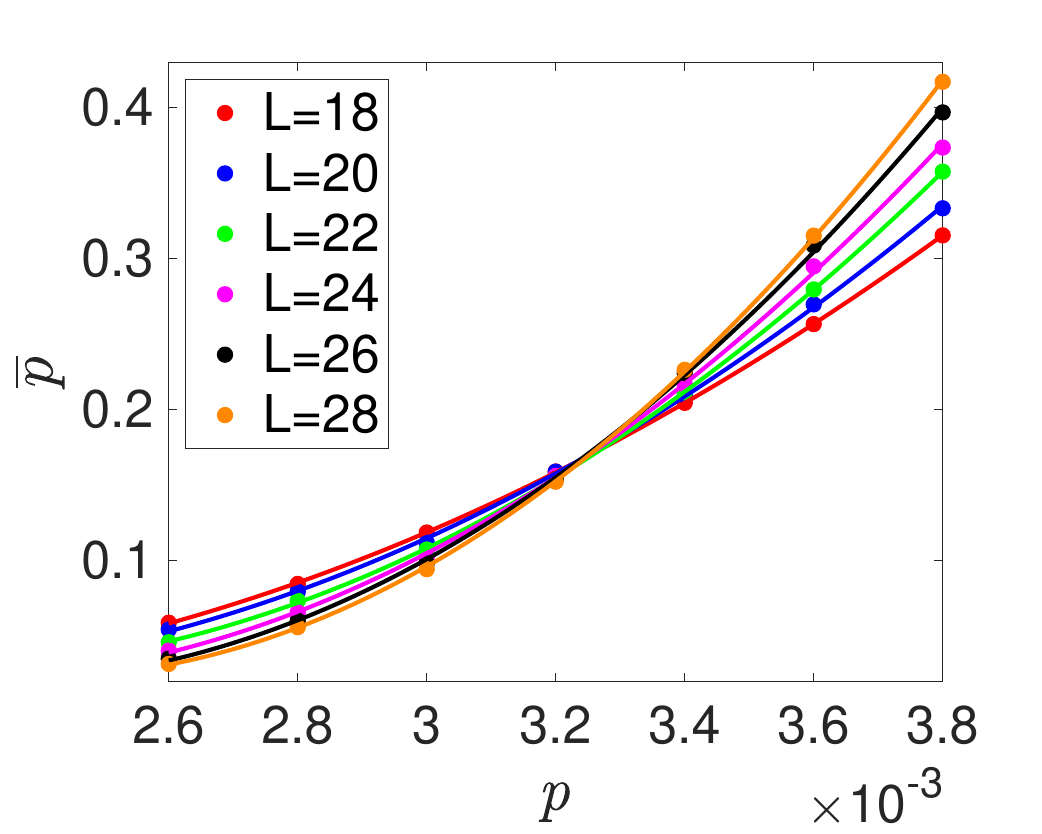}
\put(-5,75){(a)} 
\end{overpic}
\begin{overpic}[width=0.6\columnwidth]{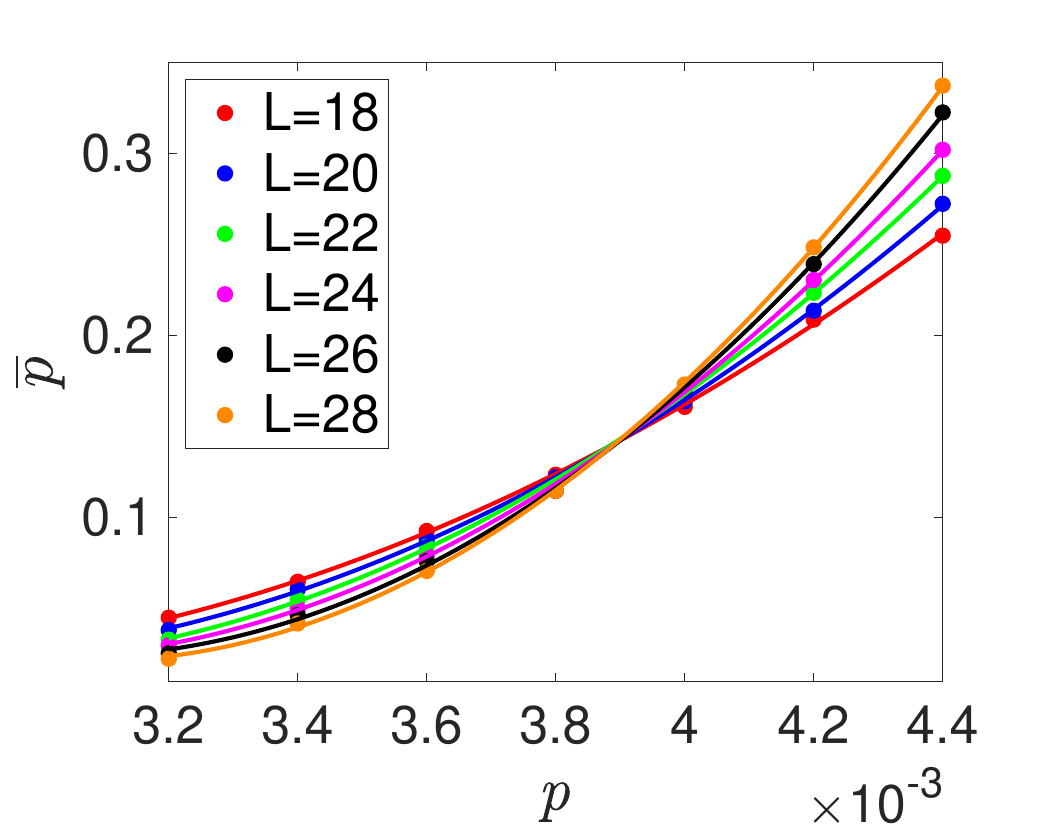}
\put(-5,75){(b)} 
\end{overpic}
\caption{Logical error rate ($\overline{p}$) versus circuit-level noise ($p$) for (a) Protocol \ref{alg1} and (b) Protocol \ref{alg2} with solid curves from fits to Eq. (\ref{eq:ansatz}). The threshold values for Protocol \ref{alg1} and \ref{alg2} extracted from the curve crossing are $0.324\%$ and $0.390\%$, respectively.}
\label{fig:3}
\end{figure}

The results of these simulations are shown in Figure \ref{fig:3}. The threshold values $p_{\rm th}$ of $0.324\%$ and $0.390\%$ were obtained for Protocols \ref{alg1} and \ref{alg2}, respectively. We remark that the threshold value of $0.390\%$ for Protocol \ref{alg2} is the same as that of Protocol B in the Ref. \cite{wan21}, as expected. However, the logical error rates of Protocol \ref{alg2} we obtain are higher than those obtained in \cite{wan21}. This is likely due to the different boundary conditions employed in our calculations. (We used the periodic boundary condition whereas Ref.~\cite{wan21} used an open boundary condition.) 

\begin{figure*}
\centering
\small
\begin{overpic}[width=0.3\textwidth]{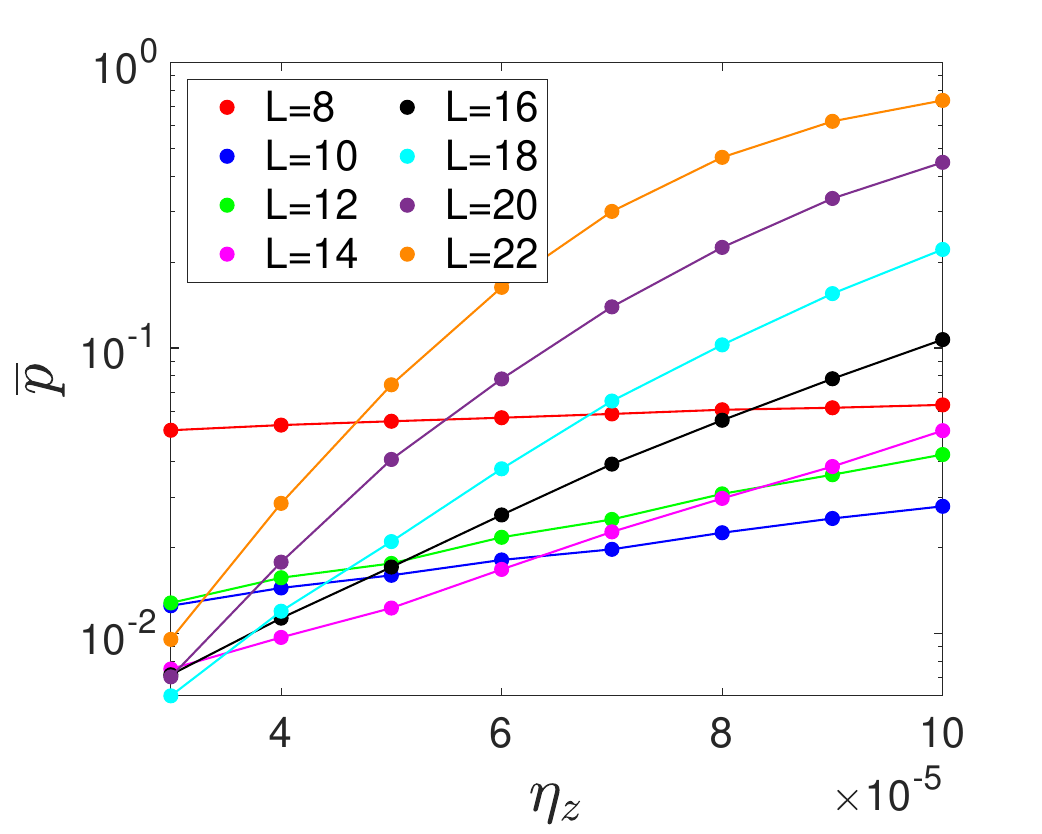}
\put(-5,75){(a)} 
\end{overpic}
\begin{overpic}[width=0.3\textwidth]{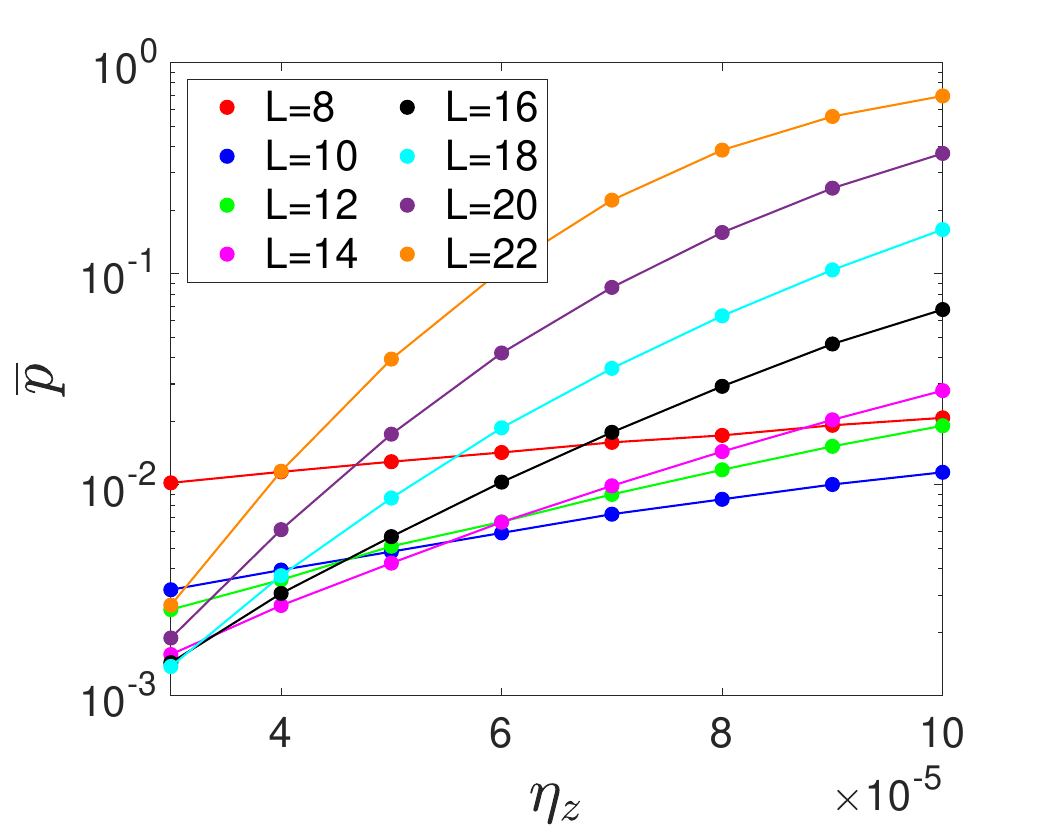}
\put(-5,75){(b)} 
\end{overpic}
\begin{overpic}[width=0.3\textwidth]{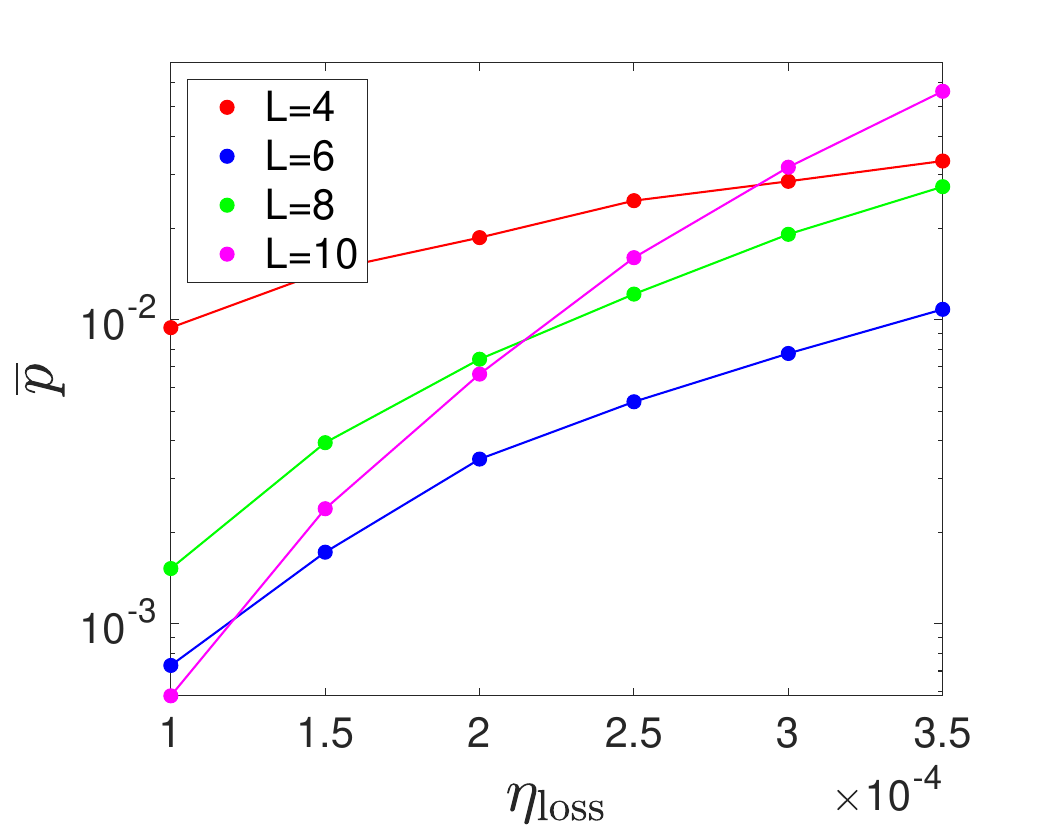}
\put(-5,75){(c)} 
\end{overpic}
\\
\begin{overpic}[width=0.3\textwidth]{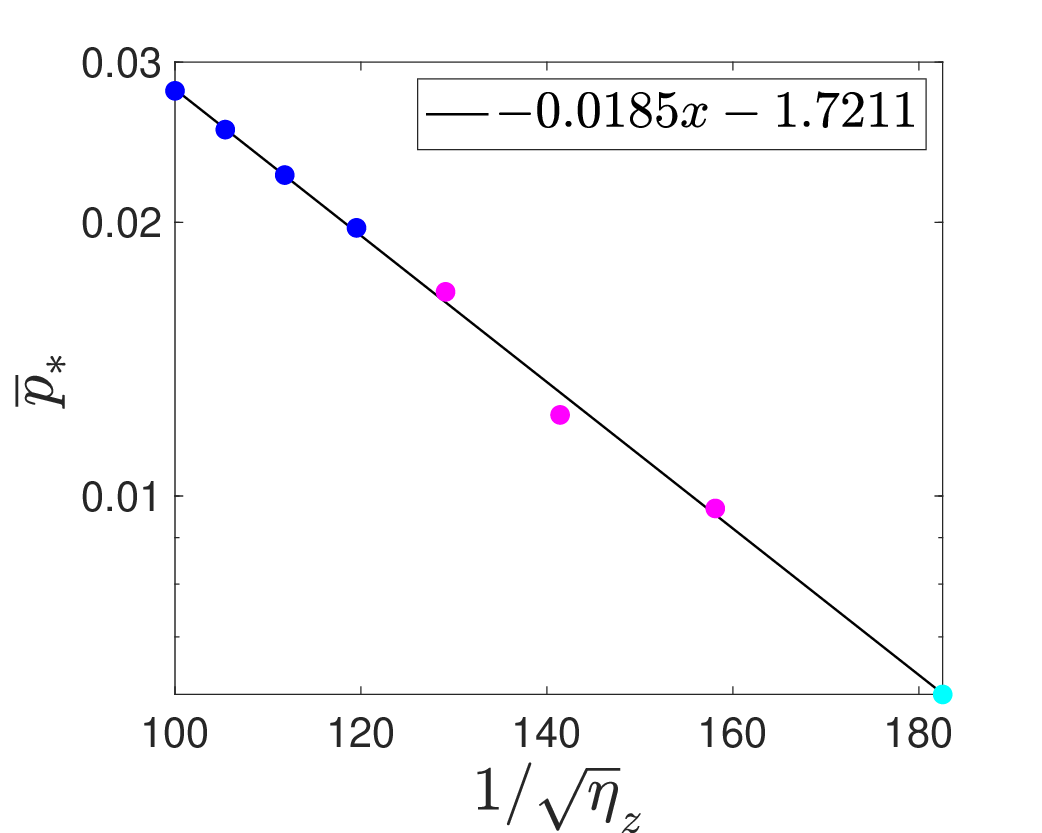}
\put(-5,75){(d)} 
\end{overpic}
\begin{overpic}[width=0.3\textwidth]{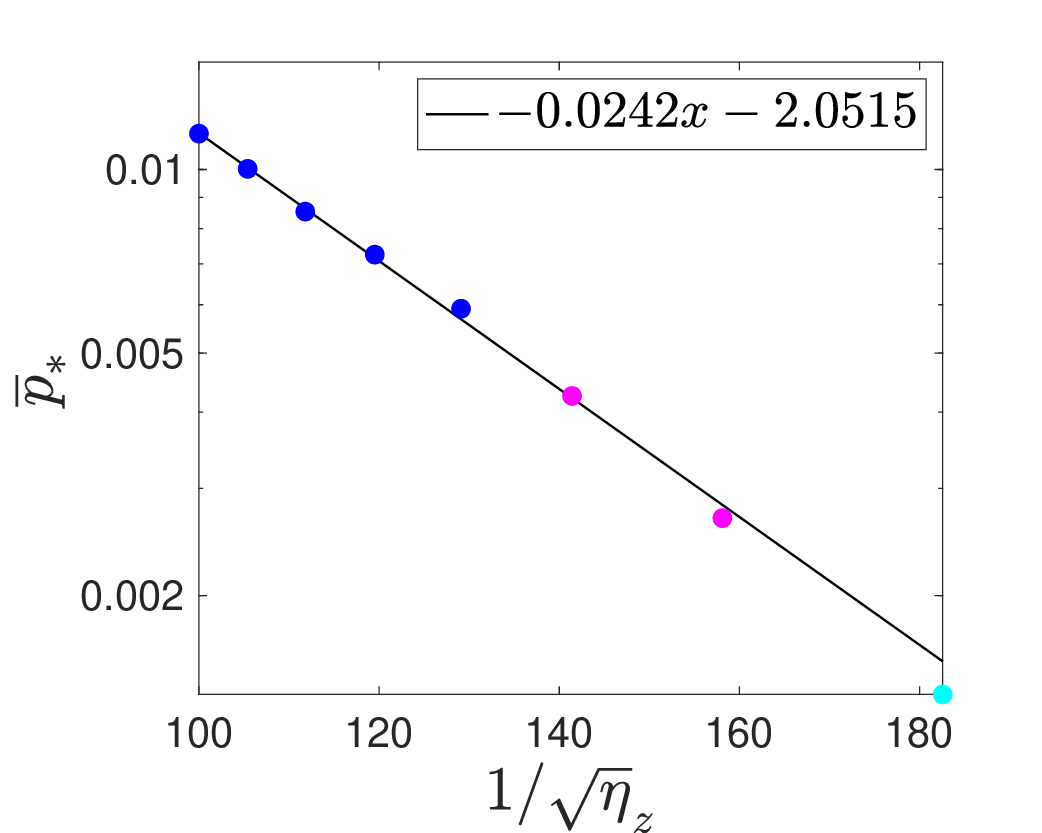}
\put(-5,75){(e)} 
\end{overpic}
\begin{overpic}[width=0.3\textwidth]{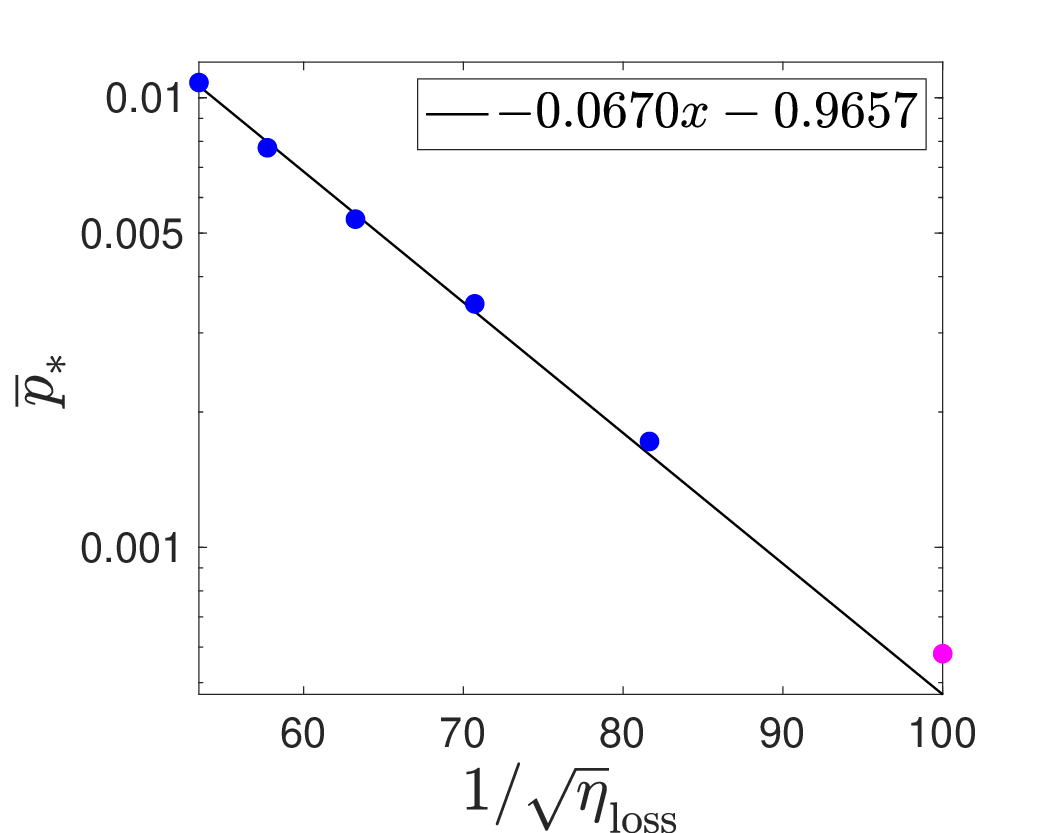}
\put(-5,75){(f)} 
\end{overpic}
\captionsetup{justification=raggedright,singlelinecheck=false}
\caption{Top panel: logical error rates $\overline{p}$ versus (a) dephasing error rate ($\eta_z$) for Protocol \ref{alg1} (b) dephasing error rates for Protocol \ref{alg2} (c) loss error rates ($\eta_{\rm loss}$). Bottom panel: $\overline{p}_*$ and its ansatz in Eq. \eqref{eq:ansatz1} versus (d) $\eta_z^{-1/2}$ for Protocol \ref{alg1} (e) $\eta_z^{-1/2}$ for Protocol \ref{alg2} (f) $\eta_{\rm loss}^{-1/2}$. The color of the dots represents the value of $L$ corresponding to the minimum $\overline{p}$ for a given $\eta_z$, with the coloring scheme following the legend in the upper figures. The $y$-axis are on a logarithmic scale.} 
\label{fig:d}
\end{figure*}

\subsubsection{Delay line error}
\label{sec:delaylineerror}
The study in Section~\ref{sec:circuit_level_noise} excludes an important source of error. Recall that the data qubits are photons, which at times propagate through a delay line. Each photon experiences an error for each unit of length they travel. We include the effect of such delay line error, following the discussion in Ref.~\cite{wan21}.

The error associated with each delay line --- from the emission of the data qubit into the delay line to the eventual measurement ---  is proportional to the time the photon spends in the delay line. The time the photon spends in the delay line is proportional to the time required to apply $U_k$, which we assume is uniform for each operation and will serve as a convenient unit of time. Throughout this paper, we will assume that there is a fixed error rate associated with this unit time, denoted as $\eta$ with an appropriate subscript, as we describe below.

There are two primary sources of errors in the delay line: dephasing and (heralded) loss error. Dephasing error is a stochastic application Pauli-$Z$ errors on the data qubit $a$ with error rate $\eta_z$:
\begin{align}
{\cal Z}_a^{(\eta_z)} (\rho)&=(1-\eta_z)\rho+\eta_z Z \rho Z .  \label{eq:etaz}
\end{align}
The errors are repeatedly applied for every unit of time to every data qubit traveling in the delay line.

For the loss error, instead of applying the loss error for every unit of time, we consider a phenomenological noise model in which a loss error proportional to the total length of the delay line is applied at the very end of the protocol. This time is proportional to $L^2$ in the leading order, and as such, the error model can be described as follows:
\begin{align}
{\cal Z}_a^{(\eta_{\rm loss})} (\rho)
=(1-L^2 \eta_{{\rm loss}})\rho+L^2\eta_{{\rm loss}} \frac{I}{2}.
\label{eq:etaloss}
\end{align}
(We will justify this phenomenological noise model further later in this Section, while discussing Table~\ref{tab:1}.)

Under the error models in Eq.~\eqref{eq:etaz} and~\eqref{eq:etaloss}, there cannot be a threshold because the error rate increases with $L$. The optimal choice of $L$ is proportional to $\eta^{-\frac{1}{2}}$ (with appropriate subscripts, depending on the error model)~\cite{wan21,Shapourian2022}, whose precise value can be obtained by increasing $L$ for a fixed value of $\eta$ until logical error rate starts to increase. 

For the readers' convenience, we briefly review the heuristic reason behind this scaling. While the strength of the circuit-level noise remains as a constant, independent of the system size, the delay line error on each qubit scales with $L^2$. If the circuit-level error is smaller than the threshold, there is a constant amount of error budget for the delay line such that, if we are below this budget, the total amount of error is still below the threshold, ensuring error suppression. This means that, insofar as $\eta L^2$ is smaller than some constant, the logical error rate is exponentially small in $L$. The best possible choice of such $L$ is clearly proportional to $\eta^{-\frac{1}{2}}$. Consequently, the heuristic estimate of optimal logical error rate (denoted as $\overline{p}_*$) is~\cite{Shapourian2022}:
\begin{align}
    \log(1/\overline{p}_*)\simeq c'\eta^{-1/2}+c'' \label{eq:ansatz1}
\end{align}
for some constants $c', c'' >0$.

Now let us discuss the simulation results for Protocols \ref{alg1} and \ref{alg2} under these error models. For the dephasing error [Eq.~\eqref{eq:etaz}], the circuit-level noise discussed in Section~\ref{sec:circuit_level_noise} remains the same. For the loss error [Eq.~\eqref{eq:etaloss}], we have removed the circuit-level noise and used the decoding algorithm in Ref.~\cite{stace10}. (As we said already, though this noise model may seem overly simplistic, we will justify this error model further.) 

For the dephasing error model [Eq.~\eqref{eq:etaz}], we averaged over $10^5$ samples for each choice of $(\eta_z,L)$, fixing the circuit-level noise as $p=10^{-3}$. For both Protocols \ref{alg1} and \ref{alg2}, Eq.~\eqref{eq:ansatz1} describes our data points well, as expected; see Figure~\ref{fig:d} (d) and (e). The same conclusion applies to the loss error model [Eq.~\eqref{eq:etaloss}] as well; see Figure~\ref{fig:d} (f).

It is interesting to see what kind of error rate ($\eta$) is needed to achieve a desired logical error rate $\overline{p}_*$. We have listed these values in Table~\ref{tab:1} (top) for the target logical error rate of $\overline{p}_*= 10^{-3}, 10^{-5}, 10^{-10}, 10^{-15}$. The same set of values was obtained in Ref.~\cite{wan21} using Protocol~\ref{alg2} and we listed them in Table~\ref{tab:1} (bottom) to compare against ours. 

Let us first compare the result obtained for the dephasing error model. One can see that, in order to achieve the same logical error rate, our scheme requires a lower dephasing error than that of Ref.~\cite{wan21}. We expect this discrepancy to be stemming from the different choice of boundary conditions; we used periodic boundary condition whereas the in Ref.~\cite{wan21} they used open boundary condition. This difference in boundary conditions results in substantially different error rates for the qubits in the first $xy$ plane, as they must wait for the construction of the last $xy$ plane to be completed under periodic boundary condition, but not under open boundary condition. This suggests that it will be more advantageous employ the open boundary condition to achieve lower logical error rate.

\begin{table}[h!]
\centering
\footnotesize
\begin{tabular}{c| c| c| c}
\hline 
\hline 
\multirow{2}{*}{\backslashbox{$\overline{p}_*$}{$\eta$}}& \multirow{2}{*}{\ref{alg1} ($\eta_z$)} & \multirow{2}{*}{\ref{alg2} ($\eta_z$)} & \multirow{2}{*}{loss ($\eta_{\rm loss}$)} \\
& & & \\
\hline
$10^{-3}$ & \SI{1.28e-5}{} & $2.48 \times 10^{-5}$ & $1.27 \times 10^{-4}$\\
\hline
$10^{-5}$ & $3.58 \times 10^{-6}$ & $6.52 \times 10^{-6}$ & $4.04\times 10^{-5}$\\
\hline
$10^{-10}$ & $7.57 \times 10^{-7}$ & $1.33 \times 10^{-6}$ & $9.23 \times 10^{-6}$\\
\hline
$10^{-15}$ & $3.19 \times 10^{-7}$ & $5.53 \times 10^{-7}$ & $3.99\times 10^{-6}$\\
\hline
\hline
\end{tabular}

\centering

\vspace*{10pt}

\begin{tabular}{c| c| c}
    \hline 
    \hline 
\multirow{2}{*}{\backslashbox{$\overline{p}_*$}{$\eta$}}& \multirow{2}{*}{dephasing ($\eta_z$)} & \multirow{2}{*}{loss ($\eta_{\rm loss}$)}\\
& & \\
\hline
$10^{-3}$ & $6.5 \times 10^{-5}$ & $7.4 \times 10^{-4}$ \\
\hline
$10^{-5}$ & $1.4 \times 10^{-5}$ & $1.4 \times 10^{-4}$ \\
\hline
$10^{-10}$ & $2.5 \times 10^{-6}$ & $2.4 \times 10^{-5}$\\
\hline
$10^{-15}$ & $1.0 \times 10^{-6}$ & $9.5 \times 10^{-6}$\\
\hline
\hline
\end{tabular}

\captionsetup{justification=raggedright,singlelinecheck=false}
\caption{(top) The requisite dephasing error rates for Protocols \ref{alg1} and \ref{alg2}, and the requisite loss error rates to achieve targeted logical error rates of $\overline{p}_*=10^{-3}, 10^{-5}, 10^{-10}, 10^{-15}$ are listed. (bottom) The same set of numbers obtained in Ref. \cite{wan21} using Protocol~\ref{alg2} is listed. These numbers are obtained by the ansatz in Eq.~\eqref{eq:ansatz1} and the results in Figure~\ref{fig:d} (d), (e), (f).
} 
\label{tab:1}
\end{table}

Now we discuss the results obtained from the loss error model. As one can see in Table~\ref{tab:1}, in order to achieve the same logical error rate, the physical loss error rate needed in our scheme is lower compared to that of Ref.~\cite{wan21}. Therefore, our estimate on the requisite loss error rate to achieve the target logical error rate can be viewed as a conservative lower bound on what is actually needed. The discrepancy between our result and the result in Ref.~\cite{wan21} is due to the choice of different boundary conditions and loss error model.

Under the assumption that the optimal choice of $L$ is proportional to $\eta^{-\frac{1}{2}}$, the requisite $L$ for the $L \times L \times L$ 3D cluster state to achieve $\overline{p}_* = 10^{-3}, 10^{-5}, 10^{-10}, 10^{-15}$ for Protocol \ref{alg1} are approximately $25$, $50$, $110$, and $170$, respectively. Similarly, the requisite $L$ for Protocol \ref{alg2} are approximately $17$, $35$, $75$, and $120$. For the last, the requisite $L$ for loss error model are approximately $7$, $13$, $30$, $45$.

Recall that the scheme in Ref.~\cite{wan21}, even applied to the state-of-the-art loss error rate delay line~\cite{Tamura2018}, does not yield any error suppression. Because the scheme discussed in this Section performs worse than that of Ref.~\cite{wan21}, this scheme will also not yield any error suppression. However, the schemes discussed in Section~\ref{sec:3} will be tantalizingly close to achieving error suppression, assuming the delay line error is equal to the one reported in Ref.~\cite{Tamura2018}; see Section~\ref{sec:4}. Accounting for the improvement in logical error rate one can have upon employing the open boundary condition (as in Ref.~\cite{wan21}), we anticipate error suppression to be well within the reach by using the multi-emitter approach delineated in Section~\ref{sec:3}, applied to the periodic boundary condition; see Section~\ref{sec:6}.

\section{Multi-emitter protocols}
\label{sec:3}

In this Section, we introduce a generalization of the protocols in Section~\ref{sec:2} to the one that involves multiple emitters. A schematic description of our setup is shown in Figure~\ref{fig:setup}.

At a high level, our approach can be explained as follows. Instead of using the single emitter to build up the entire cluster state, we distribute this task to multiple emitters and delay lines. Each emitter is connected to two delay lines as before, and the emitters form a linear array, as shown in Figure~\ref{fig:setup}. Each emitter is responsible for building up a thin ``slab'' of 3D cluster state, whose thickness is $L/n_e$ for the cluster state of size $L\times L \times L$, where $n_e$ is the number of emitters.

Of course, as it stands the state being created will be simply a set of disconnected 3D cluster states on different slabs. In order to build an isotropic 3D cluster state of size $L\times L \times L$, we need to further generate entanglement between these slabs. Happily, such entanglement can be generated by a simple modification of the single-emitter protocol [Section~\ref{sec:2}], by intermittently applying an entangling gate between the neighboring emitters without having to entangle the data qubits belonging to different slabs. 

We will first go through a simple example illustrating this idea in Section~\ref{sec:two_emitters}. The general procedure shall be explained in Section~\ref{sec:two_emitters_generalization}.

\subsection{Two emitters: an example}
\label{sec:two_emitters}

Before we describe our protocol in its full generality, let us start with an instructive example, focusing on the part of the protocol that differs the most from the single-emitter protocol [Section~\ref{sec:2}]. Without loss of generality, suppose our goal is to prepare a cluster state associated with a graph $G = (V, E)$. Assume that this graph can be partitioned into two subgraphs $G_1 = (V_1, E_1)$ and $G_2 = (V_2, E_2)$ so that (i) $V= V_1\cup V_2$ and (ii) $E_1$ and $E_2$ are edges inherited from $V_1$ and $V_2$, respectively. Note that there are additional edges connecting the vertices across $V_1$ and $V_2$; we will denote the set of these edges as $E_{12}$.

By running the single-emitter protocol for $G_1$ and $G_2$ concurrently, it is straightforward to create a tensor product of cluster states, each associated with $G_1$ and $G_2$. However, we also need to generate entanglement associated with the edges in $E_{12}$. How can we do that?

\begin{figure}[ht]
\centering
\includegraphics[width=0.35\textwidth]{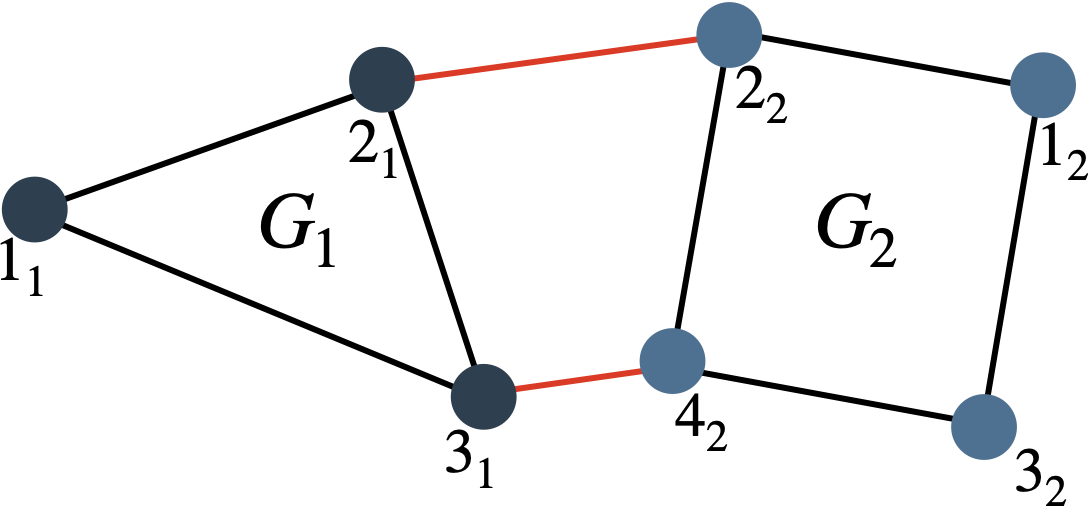}
\caption{An example for a graph $G=(V_1 \cup V_2, E_1 \cup E_2 \cup E_{12})$ where $G_1=(V_1,E_1)$ and $G_2=(V_2,E_2)$. The edges in $E_{12}$ are represented by red lines.}
\label{fig:g12}
\end{figure}

Consider an example of a graph $G$ shown in Figure~\ref{fig:g12}. If we grow the cluster state associated with the graph $G$ by adding one vertex and the edges connected to that vertex at a time following the numbering convention, there will be two steps in which we would need to generate an entanglement associated with edges in $E_{12}$: $\{2_1, 2_2 \}$ and $\{3_1, 4_2\}$. (Here the subscripts $1$ and $2$ represent the vertices in $V_1$ and $V_2$, respectively.) Because the underlying procedure is similar, we will focus on the latter. Suppose we have already created a cluster state associated with all the vertices in Figure~\ref{fig:g12} except $3_1$ and $4_2$. When we add $3_1$ and $4_2$ to the existing cluster, we need to add new edges, some of which are in $E_1\cup E_2$ and the rest in $E_{12}$. 

If we were to only add the edges in $E_{1}\cup E_2$, we can simply follow the single-emitter protocol [Section~\ref{sec:2}]. For each subgraph $G_1$ and $G_2$, we can introduce an ancilla, denoted as $Q_1$ and $Q_2$, respectively. We can apply the $\textsf{CZ}$ gate between $Q_1$ and the vertices in $V_1$ connected to $3_1$. Similarly, we can apply the $\textsf{CZ}$ gate between $Q_2$ and the vertices in $V_2$ connected to $4_2$. Then, we can apply the $\textsf{CX}$ and $\textsf{H}$ to effectively swap the state of the ancilla qubits to some data qubits [Eq.~\eqref{eq:id}], followed by a  procedure that disentangles the ancilla qubits from the rest. 

However, before we swap the states, notice that we have an opportunity to apply a \textsf{CZ} gate between the ancilla qubits. By doing so, we can generate an entanglement between the two ancilla qubits. If we then follow the rest of the procedure, e.g., swapping the state of the qubits and dientangling the ancilla qubits, we end up exactly with the cluster state we want. Therefore, the modified protocol simply involves adding a single additional \textsf{CZ} gate between the ancilla qubits prior to swapping the state of the ancilla and the data qubits. 

\begin{figure}[ht]
\centering
\includegraphics[width=0.45\textwidth]{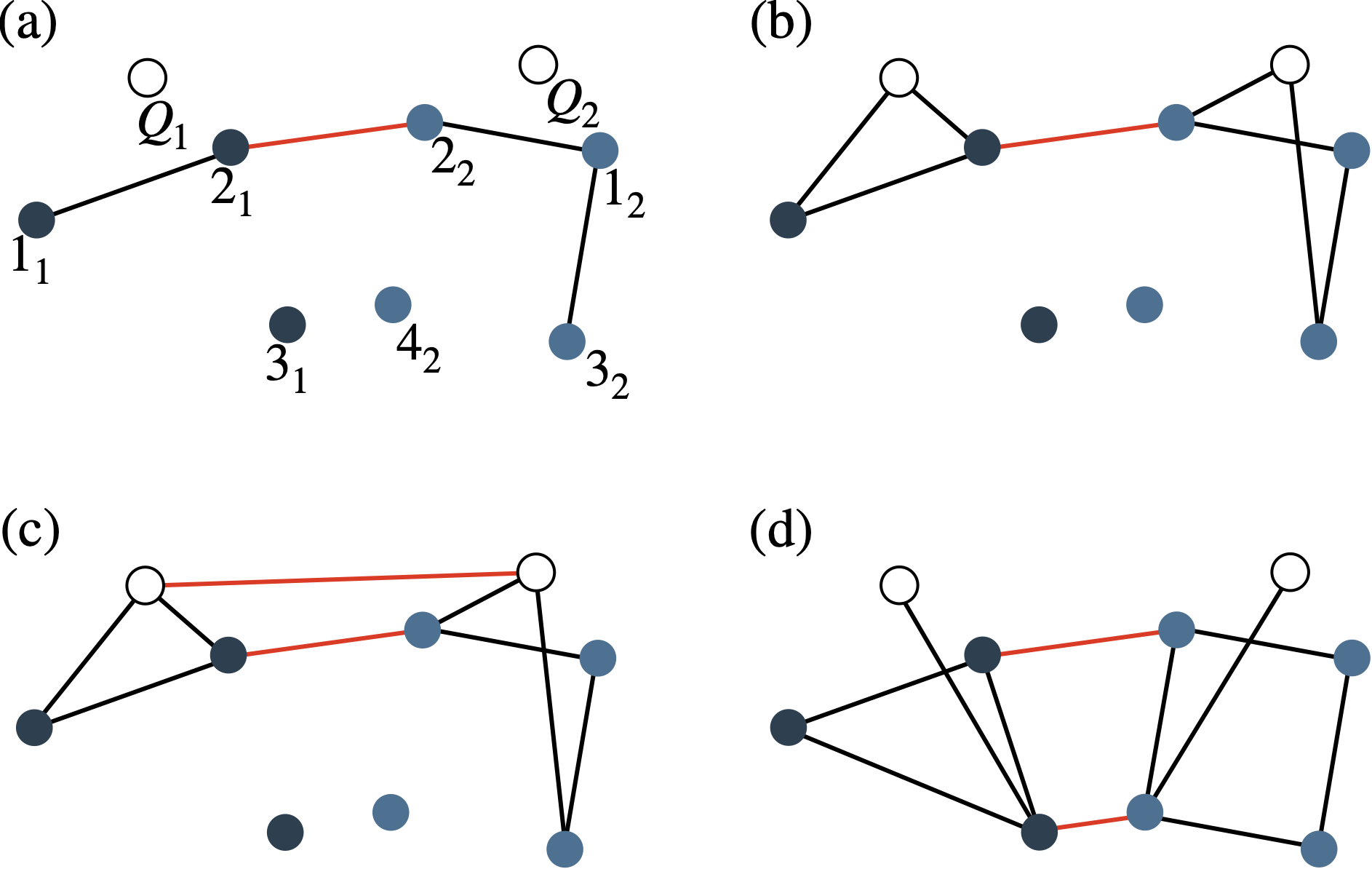}
\caption{A procedure for generating the cluster state on the graph in Figure~\ref{fig:g12} using two emitters. Two more data qubits $3_1$ and $4_2$ are incorporated into the cluster state after the procedure.
}
\label{fig:g12c}
\end{figure}

We now discuss the overall procedure in more detail. First, consider the situation where the vertices have been connected to form a cluster state except that a graph associated with all the vertices $3_1$ and $4_2$ as well as the two ancillas $Q_1$ and $Q_2$ are decoupled from the data qubits [Figure~\ref{fig:g12c} (a)]. Then we apply the \textsf{CZ} gates between $Q_1$ and $2_1$, and between $Q_2$ and $2_2$ [Figure~\ref{fig:g12c} (b)]. Next, we apply an entangling \textsf{CZ} gate between the ancilla qubits [Figure~\ref{fig:g12c} (c)]. Lastly, applying the procedure in Eq.~\eqref{eq:id}, we obtain a cluster state involving two additional data qubits $3_1$ and $4_2$ [Figure~\ref{fig:g12c} (d)]. The two ancilla qubits can be disentangled from the rest of the cluster state by either applying \textsf{CZ} gates or measuring the ancilla qubits in the $Z$-basis.

\subsection{Generalization}
\label{sec:two_emitters_generalization}

The main difference between the single-emitter protocol [Section~\ref{sec:2}] and the examplary two-emitter protocol in Section~\ref{sec:two_emitters} is the presence of an additional emitter and the \textsf{CZ} gate between the emitters. The prescription of adding such \textsf{CZ} gates generalizes straightforwardly to the generation of arbitrary cluster states, using arbitrary number of emitters, provided a certain condition is met. We discuss this generalization (as well as the condition we need) below. 

Because the multi-emitter case is a simple generalization of the two-emitter case, we focus on the protocol for two emitters. As before, we consider a partition of a graph $G=(V,E)$ into two subgraphs $G_1=(V_1,E_1)$ and $G_2= (V_2,E_2)$, so that $V= V_1\cup V_2$. The edges between $V_1$ and $V_2$ are denoted as $E_{12}$. $Q_1$ and $Q_2$ serve as ancilla qubits for $G_1$ and $G_2$, respectively.

For our protocol, we demand the following condition: each vertex in $V_1$ is connected to at most one vertex in $V_2$, and vice versa. We can then determine the appropriate numbering of data qubits for the construction of the corresponding cluster state. This condition is crucial, as the construction of the cluster state depends on interactions between $Q_1$ and the data qubits in $G_1$, between $Q_2$ and the data qubits in $G_2$, and between $Q_1$ and $Q_2$. A data qubit can be connected to multiple data qubits within the same subgraph because the ancilla qubit can interact directly with all qubits in its subgraph. However, since $Q_1$ ($Q_2$) cannot directly interact with data qubits in $G_2$ ($G_1$) and only interacts with $Q_2$ ($Q_1$), it is physically impossible for a data qubit to connect to more than one data qubit in a different subgraph.

In analogy with our discussion on the single-emitter protocol [Eq.~\eqref{eq:G-and-G'}], let us define a subgraph $G[i_1,j_2]$ and a quasi-subgraph $G[i_1,j_2]'$ of the graph $G$ as follows:
\begin{align}
G[i_1,j_2] &\equiv ([i_1]\cup[j_2] ,E_1[i_1]\cup E_2[j_2]\cup E_{12}[i_1,j_2]),\nn
G[i_1,j_2]' &\equiv ([i_1]\cup[j_2]\cup \{Q_1,Q_2\} ,E_1[i_1]\cup E_2[j_2]\cup \nn
&~~~~E_{12}[i_1,j_2] \cup \{ \{Q_1,i_1\} , \{Q_2,j_2\} \}) . 
\end{align}
Here $[i_{1(2)}] = \{1_{1(2)},\cdots, i_{1(2)}\}$ is the numbered set of data qubits in $G_{1(2)}$, and $E_{12}[i_1,j_2] = \{\{k_1,l_2\} \in E_{12}:k\in [i_1],l \in [j_2]\}$ is the set of edges connecting the two subgraphs $G_1$ and $G_2$. The quasi-subgraph $G[i_1,j_2]'$ contains, in addition to the vertices and edges in $G[i_1,j_2]$, two vertices $Q_1$, $Q_2$ and edges $\{Q_1,i_1\},~\{Q_2,j_2\}$.

The construction of the cluster state is based on a sequential generation of a cluster state associated with a quasi-subgraph $G[i_1, j_2]'$, incrementing indices ($i_1$ or $j_2$) by one at each step. Incrementation of $i_1$ and $j_2$ are mediated by the ancilla qubits $Q_1$ and $Q_2$, respectively. If neither $(i+1)_1$ nor $(j+1)_2$ are connected by an edge in $E_{12}$, we will employ the single-emitter protocol [Section \ref{sec:3}]. Since we have two ancilla qubits, both indices can be simultaneously increased by one, without affecting each other.

If $(i+1)_1$ is connected by an edge in $E_{12}$ while $(j+1)_2$ is not, we will utilize the single-emitter protocol to increment the index $j_2$. This means the incrementation of $i_1$ should be interrupted. Conversely, if $(j+1)_2$ is linked by an edge in $E_{12}$ but $(i+1)_1$ is not, we will employ the single-emitter protocol to increase the index $i_1$, and incrementing $j_2$ must be interrupted. If both $(i+1)_1$ and $(j+1)_2$ are connected by an edge in $E_{12}$, we apply the procedure analogous to the one appearing in Figure~\ref{fig:g12c}(a-d). 

This entails applying the following sequence of gates:
\begin{enumerate}
    \item We begin with a cluster state associated with a quasi-subgraph $G[i_1, j_2]'$. 
    \item In order to disentangle $Q_1$ and $Q_2$ from $i_1$ and $j_2$ respectively, we apply $Z_{Q_2,j_2}$ and $Z_{Q_1, j_1}$.
    \item By applying $\Bigl[\prod_{k_1:\{(i+1)_1,k_1\} \in E_1[i+1]} Z_{Q_1,k_1}\Bigl]$ and $\Bigl[\prod_{l_2:\{(j+1)_2,l_2\} \in E_2[j+1]} Z_{Q_2,k_2}\Bigl]$, we entangle $Q_1$ and $Q_2$ with the qubits in $V_1$ and $V_2$ that $(i+1)_1$ and $(j+2)_2$ must be entangled with.
    \item Apply $Z_{Q_1}, Q_2$ to entangle $Q_1$ and $Q_2$. 
    \item Apply $X_{Q_1, (j+1)_1}$ followed by $H_{Q_1}$, and similarly, $X_{Q_2, (j+2)_1}$ followed by $H_{Q_2}$. \label{item:last}
 \end{enumerate}
Note that the last step [Step~\ref{item:last}] realizes a swap between $Q_1$ and $(i+1)_1$ followed by a \textsf{CZ} gate between the two (and the same set of gates between $Q_2$ and $(j+1)_2$) [Eq.~\eqref{eq:id}].

Thus the overall procedure can be described as follow:
\begin{widetext}
\begin{align}
\ket{\psi_{G[(i+1)_1,(j+1)_2]'}}&=H_{Q_2} X_{Q_2,(j+1)_2} H_{Q_1} X_{Q_1,(i+1)_1}  Z_{Q_1,Q_2} \Bigl[\prod_{l_2:\{(j+1)_2,l_2\} \in E_2[j+1]} Z_{Q_2,l_2}\Bigl]   \nn
&\times \Bigl[\prod_{k_1:\{(i+1)_1,k_1\} \in E_1[i+1]} Z_{Q_1,k_1}\Bigl]Z_{Q_2,j_2} Z_{Q_1,i_1} \ket{\psi_{G[i_1,j_2]'}} \otimes \ket{0}_{(i+1)_1} \otimes \ket{0}_{(j+1)_2}.\label{eq:aaint}
\end{align}
\end{widetext}

Repeating this procedure, we can obtain the cluster state associated with the quasi-subgraph $G[|V_1|_1,|V_2|_2]'$. We can then apply $Z_{Q_1, |V_1|_1}$ and $Z_{Q_2, |V_2|_2}$ to disentangle the ancilla qubits from the data qubits, obtaining $\ket{\psi_{G}}$.

An extension of this method to the multi-emitter case is straightforward. Provided that each vertex in each subgraph is connected to at most one vertex in other subgraphs, the exact same procedure works. This means that a vertex connected to multiple vertices residing in distinct subgraphs is not allowed. Then, one can determine the appropriate numbering of data qubits for constructing the corresponding cluster state. Generalization of our approach to the one that does not require the connectivity constraint is left as a future work.

\subsection{Optimized protocol for 3D cluster state}
\label{sec:multi_3D cluster state_optimized}

Building upon the multi-emitter protocol discussed in Section~\ref{sec:two_emitters_generalization}, we now discuss the protocols tailored to creating the 3D cluster state, called \Mref{alg3} and \Mref{alg4}.\footnote{We stated these protocols under the assumption that $|V_1|=|V_2|$ and all edges in $E_{12}$ are represented as $\{i_1,i_2\}$.} (Here M stands for the \emph{multi}-emitter protocol.) We optimized this protocol (compared to the one discussed in Section~\ref{sec:two_emitters_generalization}) by removing the redundant gates; see the \textbf{if} statements therein. The operations $O_{i_{1(2)}}$ and $O_{i_{1(2)}}'$ are defined as:
\begin{widetext}
\begin{align}
O_{i_{1(2)}}&= \Bigl[\prod_{j_{1(2)}:\{i_{1(2)},j_{1(2)}\} \in E_{1(2)}[i]} Z_{Q_{1(2)},j_{1(2)}}\Bigl]Z_{Q_{1(2)},i_{1(2)}}\nn
O_{i_{1(2)}}'&= \Bigl[\prod_{j_{1(2)}\neq (i-1)_{1(2)} \& j_{1(2)}:\{i_{1(2)},j_{1(2)}\} \in E_{1(2)}[i]} Z_{Q_{1(2)},j_{1(2)}}\Bigl].
\end{align}    
\end{widetext}

The main difference between the two protocols is the way in which we implement $U_k$ for vertices that are not connected to vertices in the other subgraphs; \Mref{alg3} uses the one in Protocol~\ref{alg1} while \Mref{alg4} uses the one in Protocol~\ref{alg2}. 

\begin{figure}[ht]
\centering
\small
\begin{overpic}[width=0.35\textwidth]{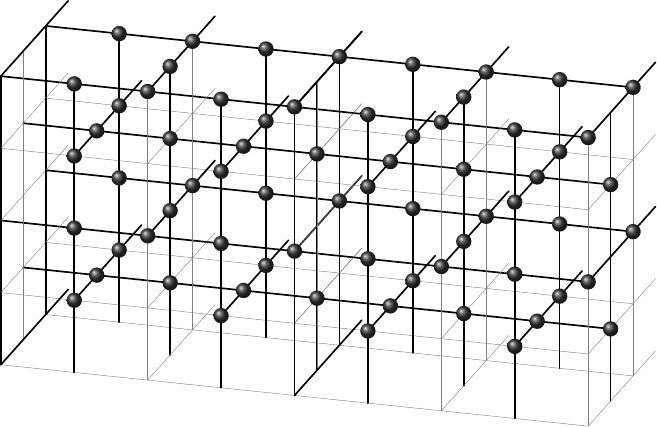}
\put(-17,64){(a)} 
\end{overpic}

\begin{overpic}[width=0.47\textwidth]{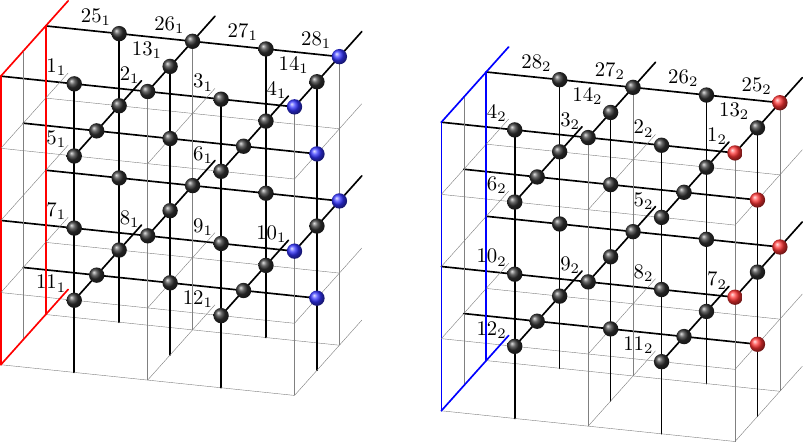}
\put(0,55){(b)} 
\end{overpic}
\caption{(a) A part of the $8 \times 8 \times 8$ 3D cluster state under periodic boundary condition is depicted. The thick edges between two qubits represent the entanglement between the two qubits. (b) The subgraphs $G_1$ and $G_2$ of the 3D cluster state are represented. The entanglement between two different subgraphs are needed for $\{1_1,1_2\},~ \{4_1,4_2\},~ \{7_1,7_2\},~ \{10_1,10_2\} \cdots$.}
\label{fig:4}
\end{figure}

\begin{figure*}[ht]
\begin{minipage}{\textwidth}
\begin{algorithm}[H]
\renewcommand{\thealgorithm}{M1, M2}
\caption{Cluster state construction with two ancilla qubits }
\small
\begin{algorithmic}[1]
\State initialize both $Q_1$ and $Q_2$ in $\ket{+}$\label{alg3}
\For{$i=1$ to $|V_1|$}\label{alg4}
    \State initialize qubits $i_1$ and $i_2$ in $\ket{0}$
    \If{$\{i_1,i_2\} \notin E_{12}$}
        \State apply the $U_k$ in Protocol \ref{alg1} to $\ket{\psi_{G[(i-1)_1,(i-1)_2]'}}$ using $Q_1$ and $Q_2$ \textit{// Protocol M1}
        \Statex ~~~~~~~~~or
        \Statex ~~~~~~~~~apply the $U_k$ in Protocol \ref{alg2} to $\ket{\psi_{G[(i-1)_1,(i-1)_2]'}}$ using $Q_1$ and $Q_2$ \textit{// Protocol M2}
    \Else
        \If{$\{(i-1)_1,i_1\} \in E_1$}
            \If{$\{(i-1)_2,i_2\} \in E_2$}
            \State apply $H_{Q_2} X_{Q_2,i_2} H_{Q_1} X_{Q_1,i_1}  Z_{Q_1,Q_2} O_{i_2}' O_{i_1}'$
            \Else
            \State apply $H_{Q_2} X_{Q_2,i_2} H_{Q_1} X_{Q_1,i_1}  Z_{Q_1,Q_2} O_{i_2} O_{i_1}'$
            \EndIf
        \Else
            \If{$\{(i-1)_2,i_2\} \in E_2$}
            \State apply $H_{Q_2} X_{Q_2,i_2} H_{Q_1} X_{Q_1,i_1}  Z_{Q_1,Q_2} O_{i_2}' O_{i_1}$
            \Else
            \State apply $H_{Q_2} X_{Q_2,i_2} H_{Q_1} X_{Q_1,i_1}  Z_{Q_1,Q_2} O_{i_2} O_{i_1}$
            \EndIf
        \EndIf
    \EndIf
\EndFor
\State apply $Z_{Q_1,|V_1|_1}$ and $Z_{Q_2,|V_2|_2}$

\end{algorithmic}
\end{algorithm}
\end{minipage}
\end{figure*}

As an example, an explicit partition of the graph associated to the 3D cluster state (under periodic boundary condition) is shown in Figure~\ref{fig:4} (b). From original 3D cluster state depicted in Figure~\ref{fig:4} (a), $G_1$ and $G_2$ corresponds to the left and the right graph in Figure~\ref{fig:4} (b) and the blue (red) data qubits correspond to the blue (red) face on the other graph. Note that each vertex in $G_1$ is connected to at most one vertex in $G_2$ and vice versa, as we demanded.

We chose the number of qubits in $G_1$ and $G_2$ to be the same and labeled the vertices in such a way that the edges in $E_{12}$ are represented by pairs of qubits labeled by the same integer (e.g. $\{1_1,1_2\}, \{4_1,4_2\} \cdots \in E_{12}$). This implies that when beginning with the cluster state associated with $G[0_1,0_2]'$ and considering the construction procedure [Section~\ref{sec:two_emitters_generalization}], two newly introduced vertices are either both unconnected by an edge in $E_{12}$ or both connected by an edge in $E_{12}$. Therefore, we can achieve a simplified delay line structure and optimize construction time, as the entire construction procedure proceeds without interruption.

We can generalize the conditions for optimizing the 3D cluster state construction procedure using $n_e$ emitters, which does not allow for interruption. Firstly, considering the fact that two emitters should interact to entangle the qubits in two different subgraphs, we need an even number of emitters; if not, an incrementation of one index for the subgraph has to be interrupted. Secondly, $L/n_e$ must be an integer to ensure that the number of qubits in different slabs is the same. In Section~\ref{sec:4}, we simulated all the 3D cluster state construction procedure satisfying these two conditions.

\section{Fault-tolerant error correction using multiple emitters}
\label{sec:4}

In this Section, we report the results of a numerical simulation of~\Mref{alg3} and~\Mref{alg4}.

\subsection{Threshold}
\label{sec:4a}

In this Section, we study the thresholds for Protocols \Mref{alg3} and \Mref{alg4}. Note that the main difference between these protocols and the single-emitter protocols [Section~\ref{sec:2}] is the presence of \textsf{CZ} gate between the ancilla qubits. We will employ the standard depolarizing noise model for this gate as well [Eq.~\eqref{eq:dep}].


The physical mechanisms governing the entangling gate between emitters may differ from those underlying the gate between an emitter and a photon~\cite{mirhosseini2019cavity,ferreira2020collapserevivalartificialatom,ferreira22,QDotQED, tiranov2023collective}. Consequently, these distinct mechanisms are generally expected to exhibit different error rates compared to the emitter-photon gate. For this reason, we tune the ratio between the two sources of error, probing the regime in which the error rate between the emitters is larger than the rest. More precisely, we denote the noise strength of the gate between the emitters as $p_e$ and the noise strength of the rest as $p$ [Eq.~\eqref{eq:dep}]. We studied the threshold under different ratios of $p_e/p$, ranging from $1$ to $10$.

\begin{figure}[h!]
\centering
\small
\begin{overpic}[width=0.3\textwidth]{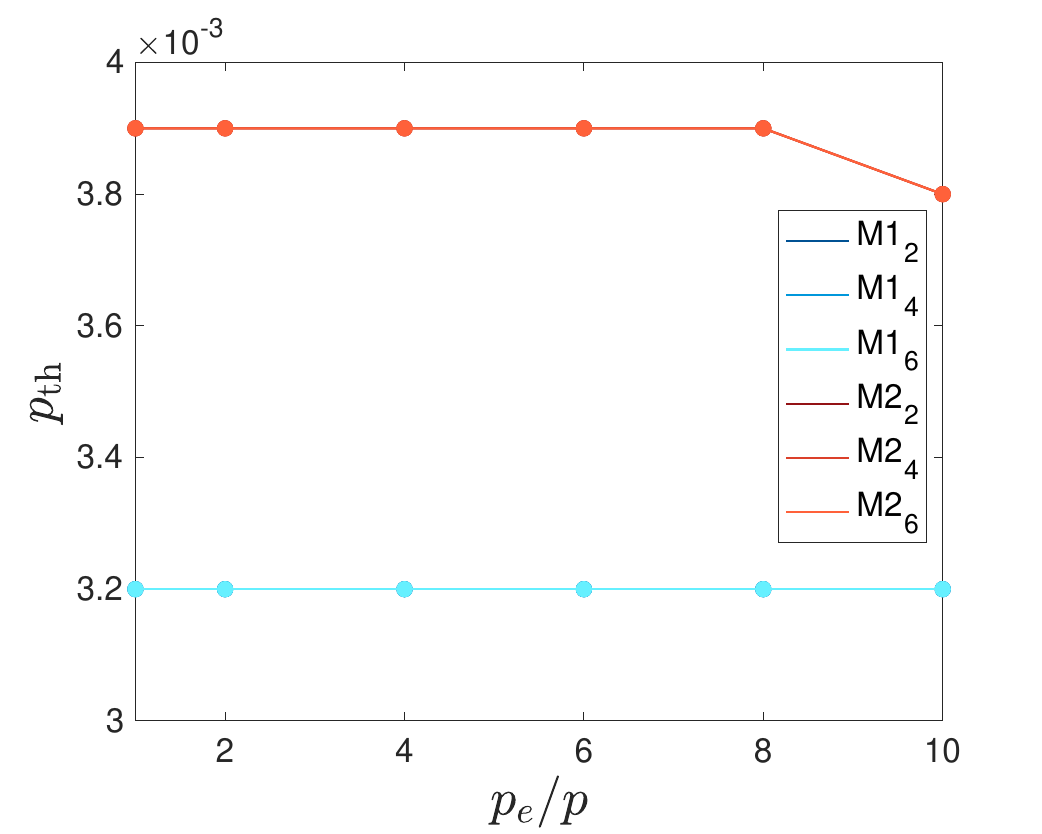}
\put(-5,75){(a)} 
\end{overpic}
\begin{overpic}[width=0.3\textwidth]{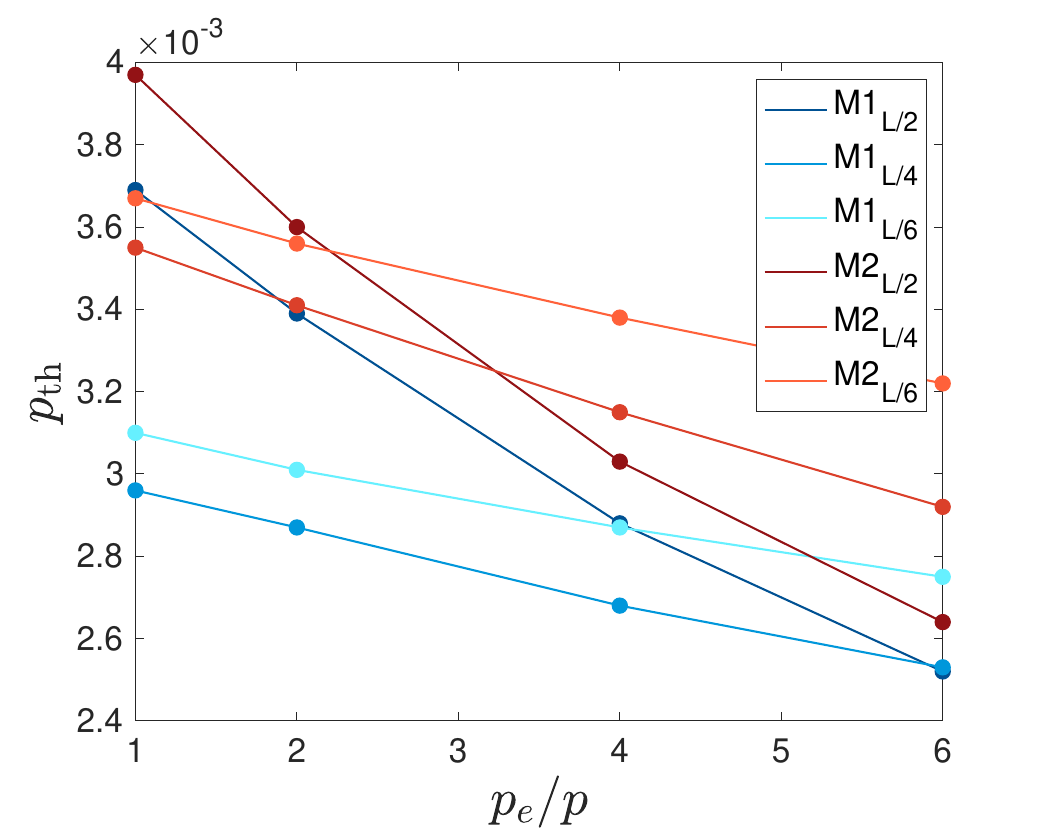}
\put(-5,75){(b)} 
\end{overpic}
\caption{(a) threshold values $p_{\rm th}$ for $p$ versus $p_e/p$ for $2,4,6$ emitters are depicted. The threshold values of \Mref{alg3}$_2$ and \Mref{alg3}$_4$ are the same as those of \Mref{alg3}$_6$, while the threshold values of \Mref{alg4}$_2$ and \Mref{alg4}$_4$ are the same as those of \Mref{alg4}$_6$. (b) threshold values $p_{\rm th}$ for $p$ versus $p_e/p$ for $L/2,L/4,L/6$ emitters are depicted. 
In the legends, the subscripts denote the number of emitters.
}
\label{fig:LL}
\end{figure}

Here are the details of our numerical study. We chose a range of values of $1 \leq p_e/p \leq 10$ for different number of emitters, using Protocol \Mref{alg3} and \Mref{alg4}. Each data point is averaged over $10^5$ samples. The thresholds were obtained by fitting the data to the ansatz in Eq. \eqref{eq:ansatz}. The results are shown in Figure~\ref{fig:LL}. Figure~\ref{fig:LL}(a) shows the threshold when the number of emitters is a constant independent of the code distance. Figure~\ref{fig:LL}(b) shows the threshold when the number of emitters scales linearly with the code distance. 

\begin{figure*}
\centering
\small
\begin{overpic}[width=0.3\textwidth]{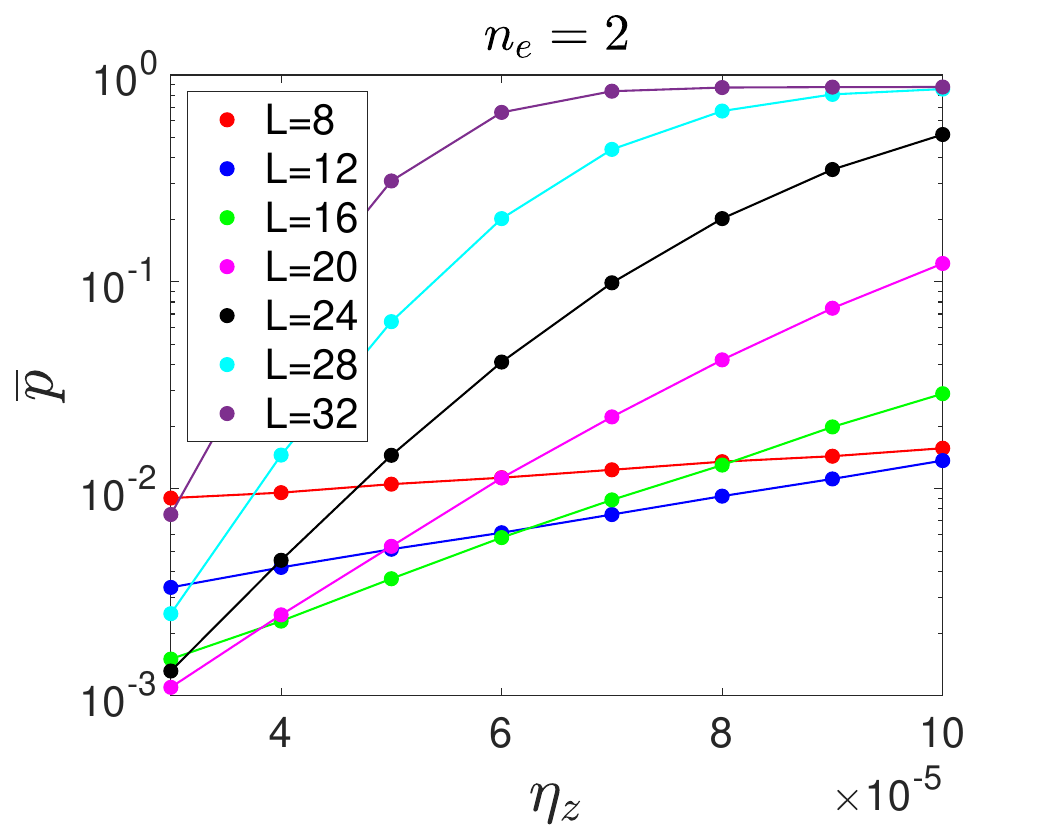}
\put(-5,75){(a)} 
\end{overpic}
\begin{overpic}[width=0.3\textwidth]{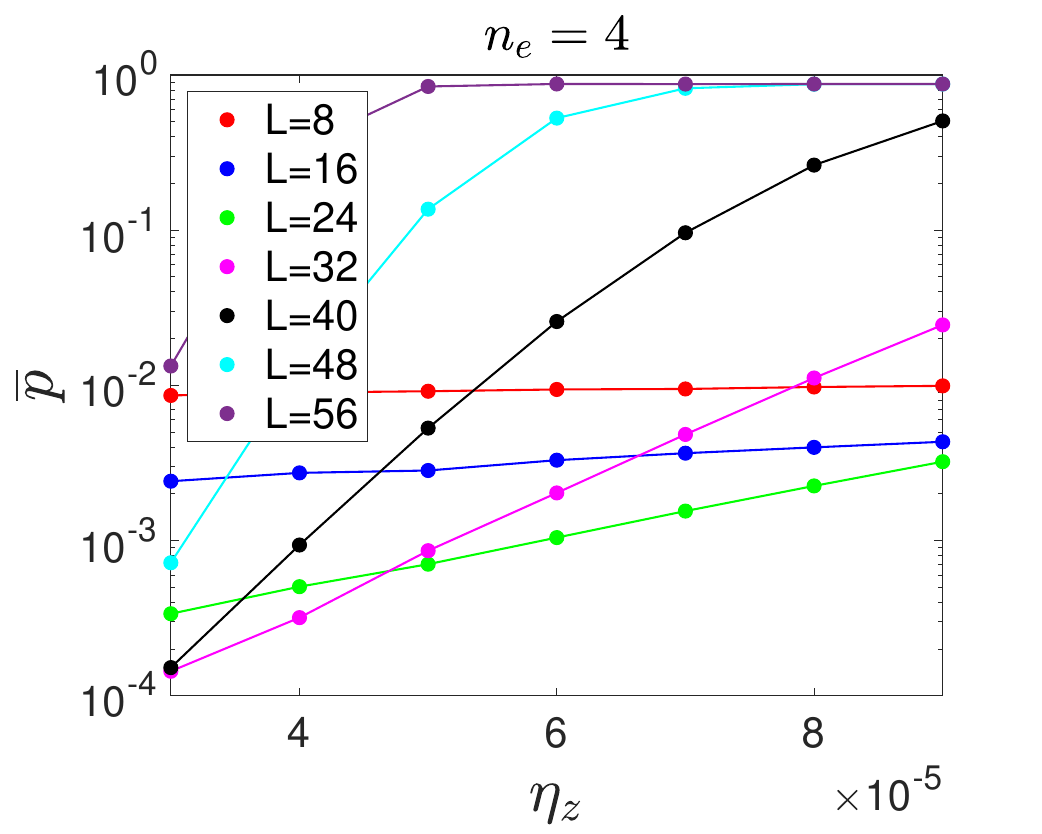}
\put(-5,75){(b)} 
\end{overpic}
\begin{overpic}[width=0.3\textwidth]{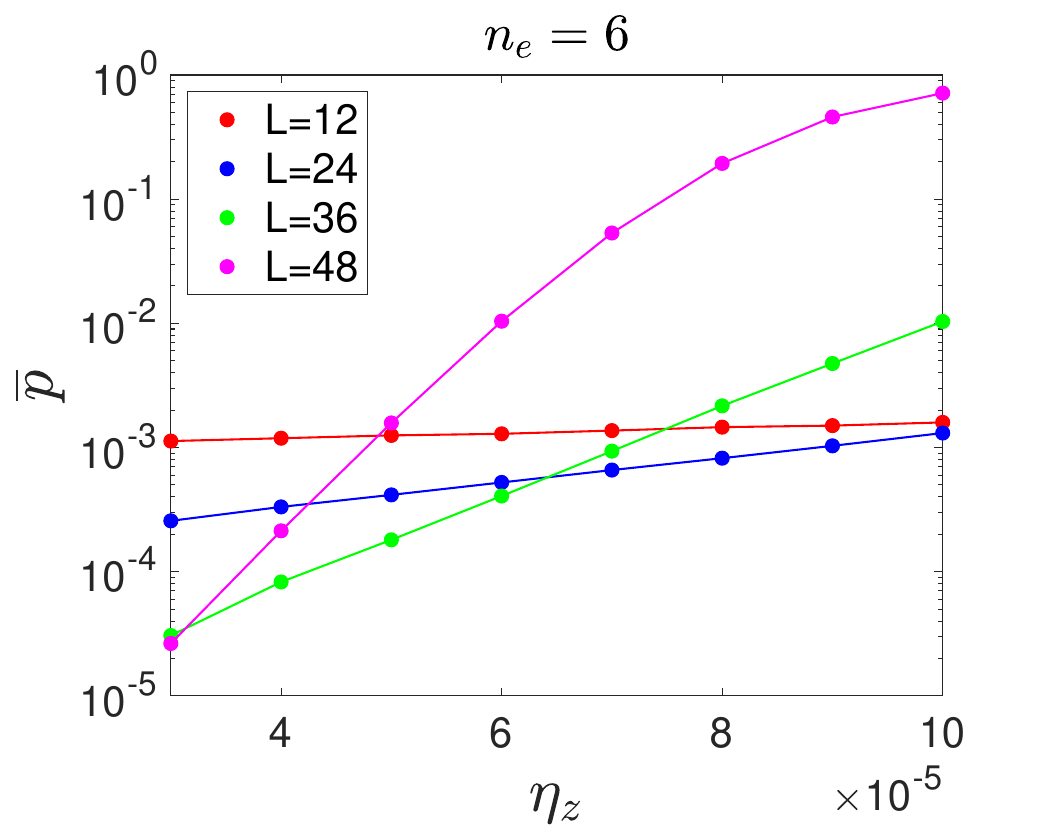}
\put(-5,75){(c)} 
\end{overpic}
\\
\begin{overpic}[width=0.3\textwidth]{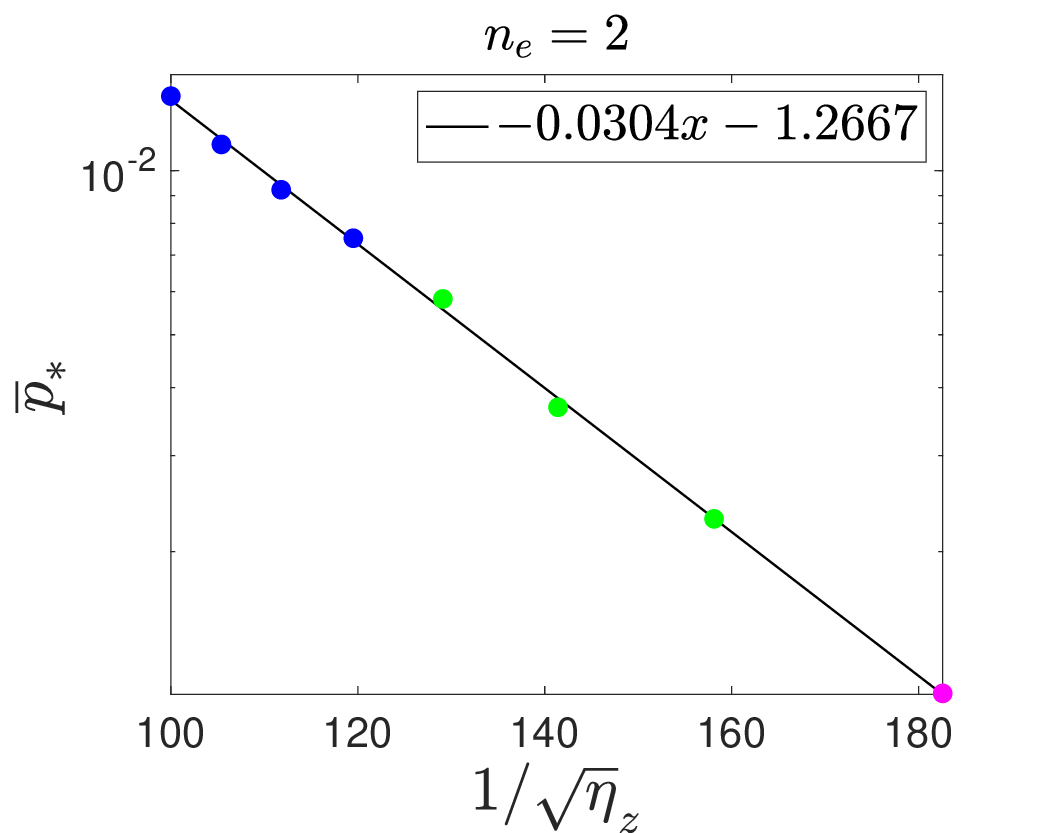}
\put(-5,75){(d)} 
\end{overpic}
\begin{overpic}[width=0.3\textwidth]{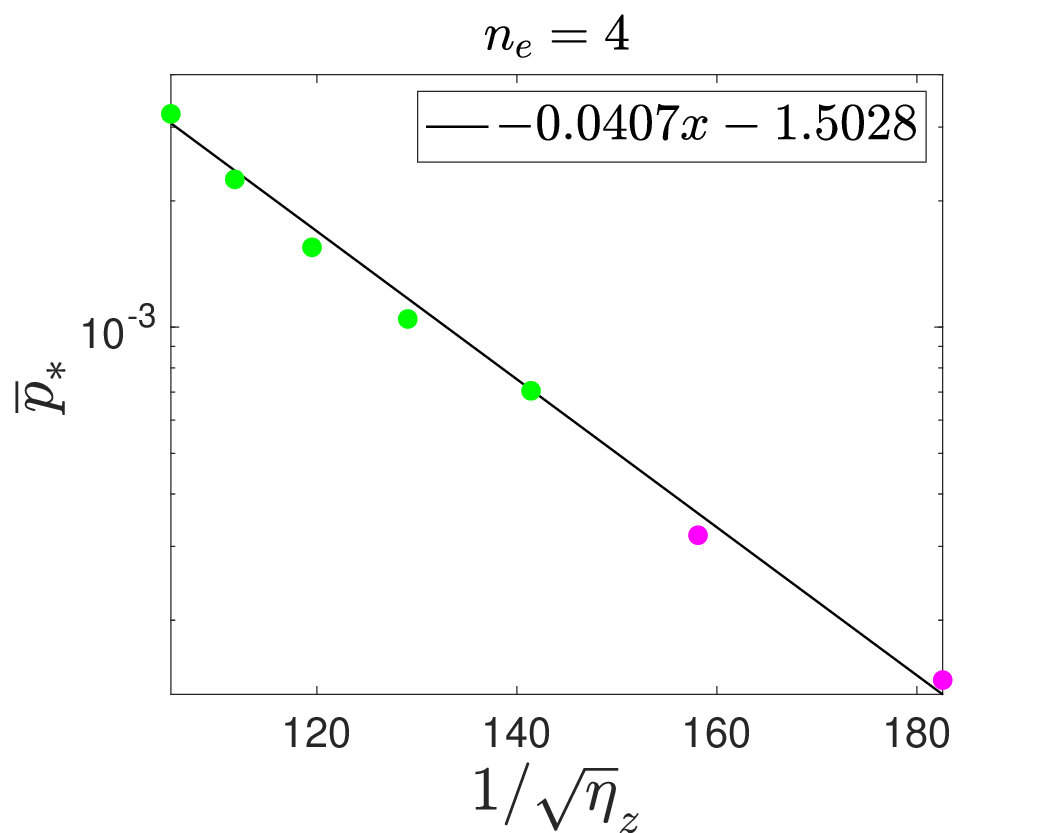}
\put(-5,75){(e)} 
\end{overpic}
\begin{overpic}[width=0.3\textwidth]{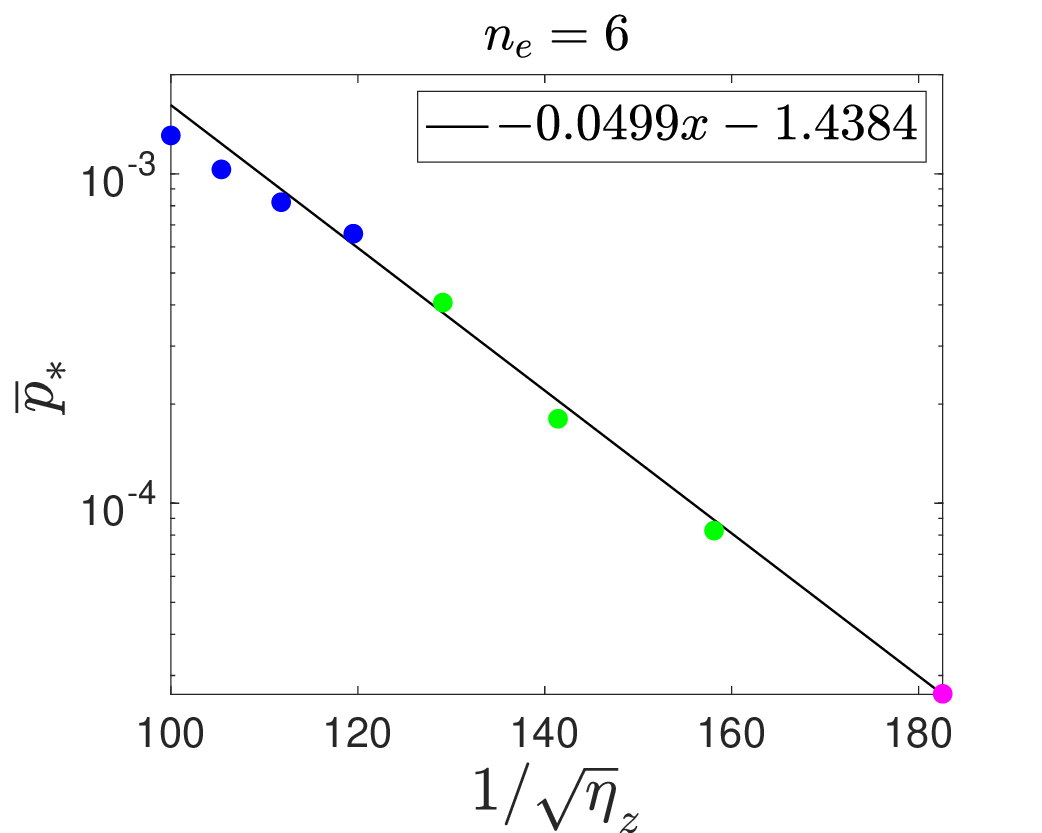}
\put(-5,75){(f)} 
\end{overpic}
\caption{Top panel: logical error rates versus dephasing error rates for (a) $2$ emitters (b) $4$ emitters (c) $6$ emitters, employing Protocol \Mref{alg3}; the numbers in the subscript represent the number of emitters. Bottom panel: $\overline{p}_*$ and its ansatz in Eq. \eqref{eq:ansatzn} versus $\eta_z^{-1/2}$ for (d) $2$ emitters (e) $4$ emitters (f) $6$ emitters, employing Protocol \Mref{alg3}. The color of the dots represents the value of $L$ corresponding to the minimum $\overline{p}$ for a given $\eta_z$, with the coloring scheme following the legend in the upper figures. The $y$-axis are on a logarithmic scale.}
\label{fig:7}
\end{figure*}

The most striking conclusion from Figure~\ref{fig:LL} is that the threshold value barely changes even if the ratio $p_e/p$ approaches $10$, provided that $n_e$ is a constant independent of the code distance [Figure~\ref{fig:LL}(a)]. This suggests that our scheme can tolerate a significantly higher error rate for the entangling gate between the emitters compared to the other sources of error. Our result is consistent with a recent study which also demonstrated a high tolerance against an error at the interface in a modular approach to fault-tolerantly correcting errors using the surface code~\cite{Ramette2023}. While the underlying error model is not identical, the analytical arguments in Ref.~\cite{Ramette2023} can explain our observation qualitatively.

In fact, there is intuitive explanation for this phenomenon. Roughly speaking, the gates between the emitters are responsible for creating the entanglement between different slabs created from each emitter. When the number of emitters is small, majority of the gates used are the gates between the emitters and photons, not between the emitters. More precisely, the number of emitter-emitter gates scales linearly with $L^2$ and the number of emitters. This is a much smaller number than the total number of gates, which scales linearly with $L^3$.

When the number of emitters scales linearly with the code distance, we do observe a decrease in threshold as $p_e/p$ increases. Nonetheless, the order of magnitude of the threshold remains the same.  

From Figure~\ref{fig:LL}, we can infer that the entanglement operation between the two ancilla qubits is unlikely to result in adverse error propagation. The reason is that, although the number of entanglement operations between the two ancilla qubits increases to construct the 3D cluster state for larger $n_e$, the threshold value of the logical error rate remains relatively stable, suggesting that the overall error performance is minimally affected.

\begin{figure*}
\centering
\small
\begin{overpic}[width=0.3\textwidth]{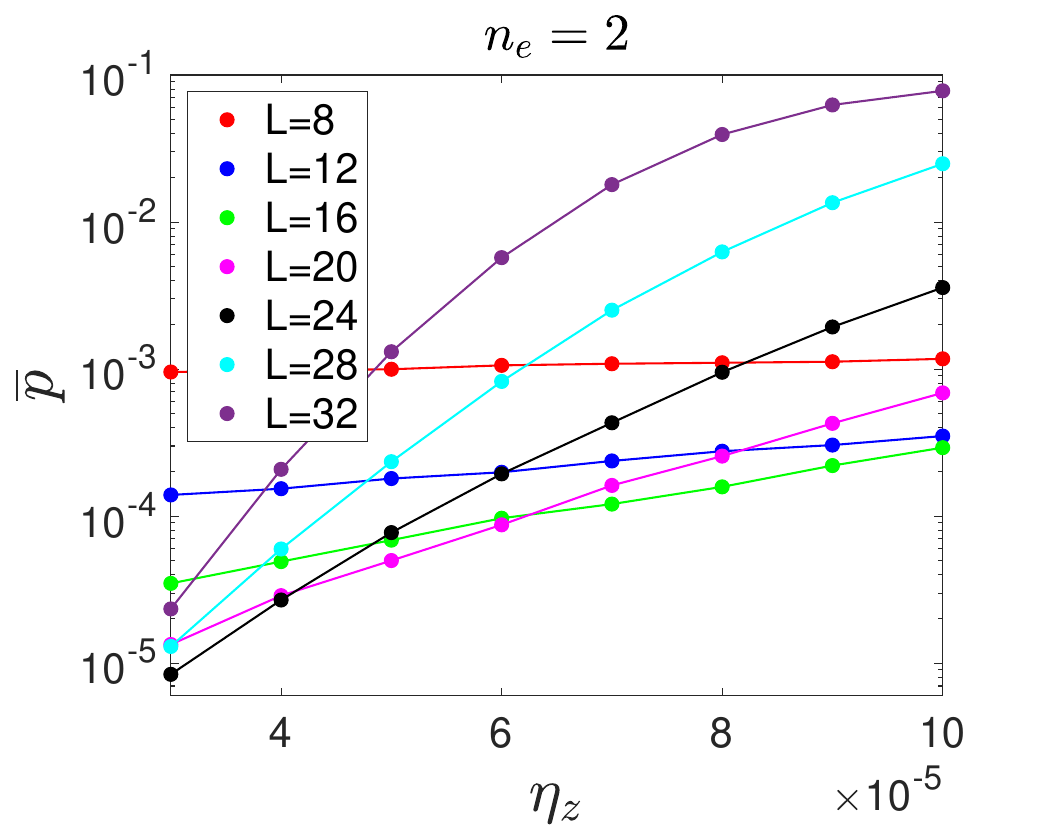}
\put(-5,75){(a)} 
\end{overpic}
\begin{overpic}[width=0.3\textwidth]{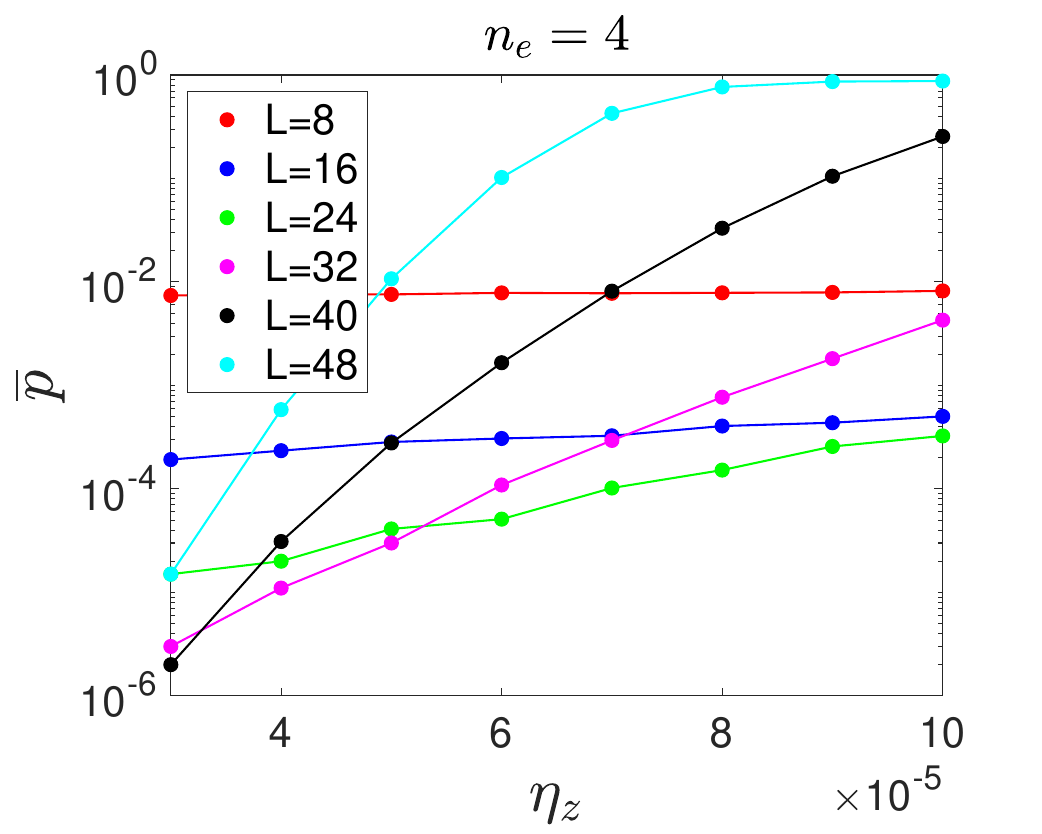}
\put(-5,75){(b)} 
\end{overpic}
\begin{overpic}[width=0.3\textwidth]{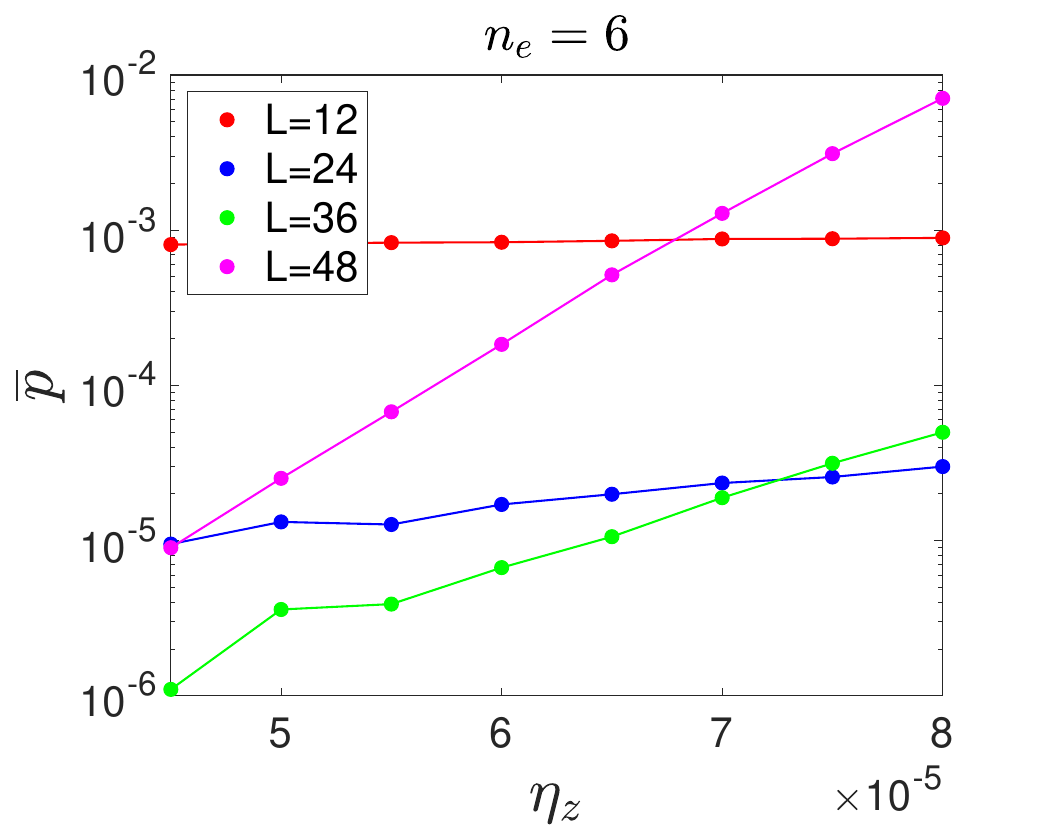}
\put(-5,75){(c)} 
\end{overpic}
\\
\begin{overpic}[width=0.3\textwidth]{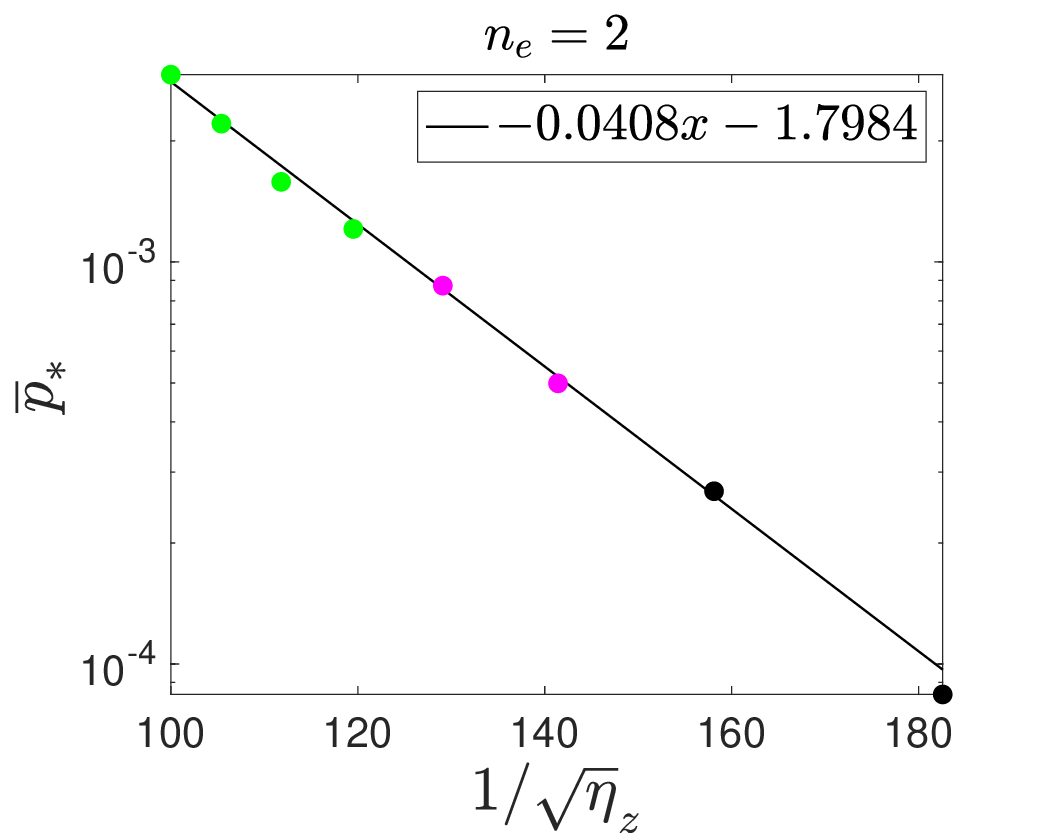}
\put(-5,75){(d)} 
\end{overpic}
\begin{overpic}[width=0.3\textwidth]{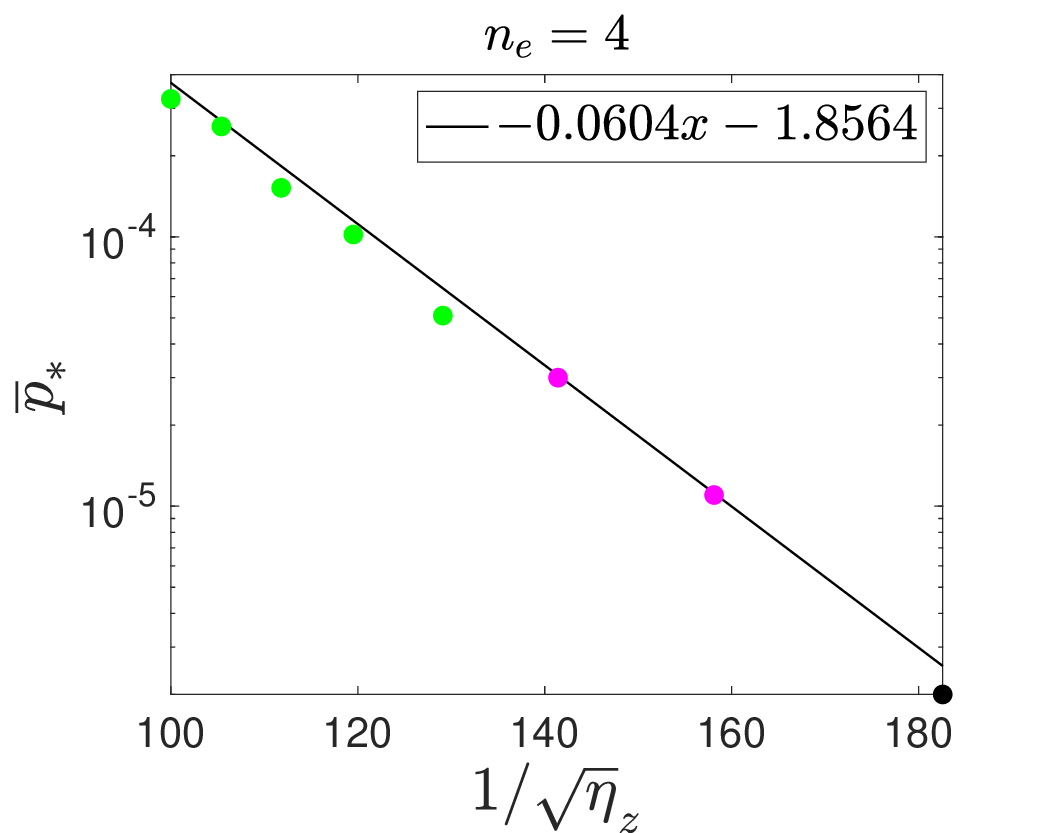}
\put(-5,75){(e)} 
\end{overpic}
\begin{overpic}[width=0.3\textwidth]{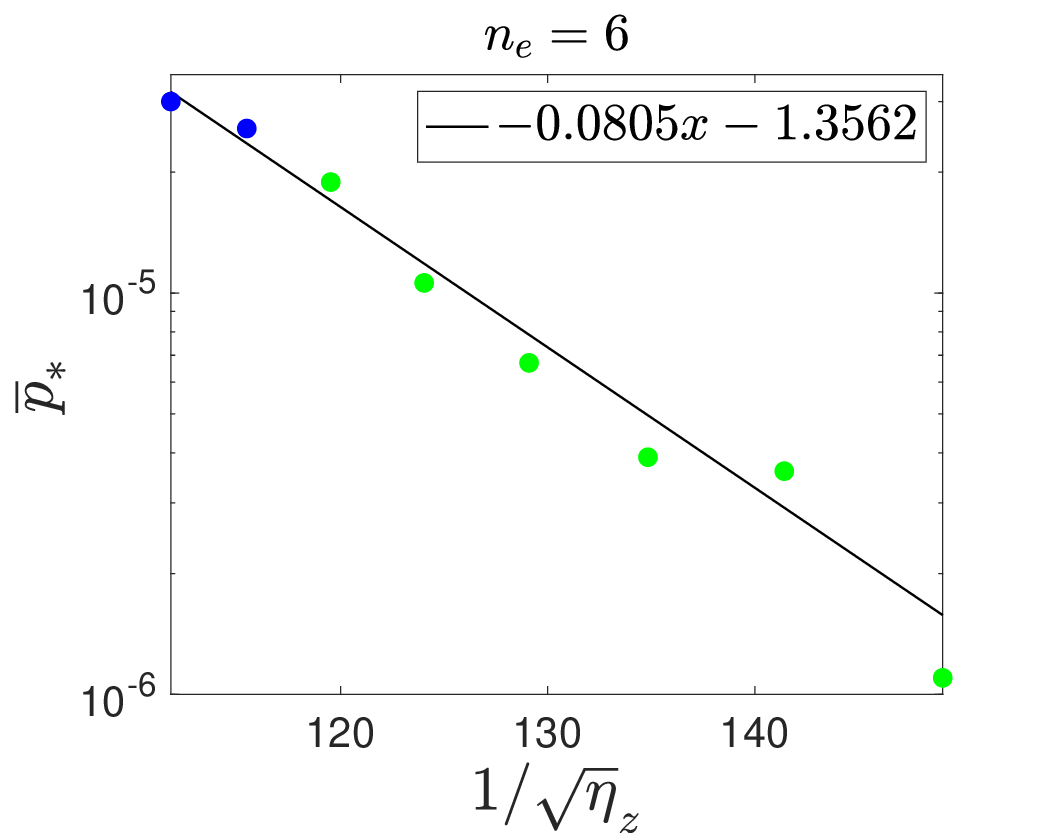}
\put(-5,75){(f)} 
\end{overpic}
\caption{Top panel: logical error rates versus dephasing error rates for (a) $2$ emitters (b) $4$ emitters (c) $6$ emitters, employing Protocol \Mref{alg4}. Bottom panel: $\overline{p}_*$ and its ansatz in Eq. \eqref{eq:ansatzn} versus $\eta_z^{-1/2}$ for (d) $2$ emitters (e) $4$ emitters (f) $6$ emitters, employing Protocol \Mref{alg4}. The color of the dots represents the value of $L$ corresponding to the minimum $\overline{p}$ for a given $\eta_z$, with the coloring scheme following the legend in the upper figures. The $y$-axis are on a logarithmic scale.}
\label{fig:8}
\end{figure*}

\subsection{Delay line error}
\label{sec:5}

We now include the effect of the delay line errors. The error model associated with this error is identical to the one used in Section~\ref{sec:delaylineerror}. For the \textsf{CZ} gate between the ancilla qubits, we assume the error rate is equal to the error rate of the other circuit-level noises, i.e., $p_e = p$.

As we discussed already in Section~\ref{sec:2}, in the presence of a delay line error, there is no threshold. A more relevant question was how one can optimally choose the length of the delay line, given the error rate per unit length $\eta$, and what the corresponding optimal logical rate $\overline{p}_*$ is. 

Here, we aim to understand the same question but instead focusing on how to obtain the logical error rate in terms of $n_e$ and $\eta$. There are two regimes of interest, depending on whether the number of emitters is small or large, compared to $L$.

For a fixed $n_e$, the strength of the delay line error on each qubit scales linearly with $\eta L^2/n_e$. Therefore, in order to ensure that the amount of accumulated error is smaller than the threshold (which is necessary for error suppression), the maximum length one can have is $L\propto \sqrt{n_e/\eta}$. Recalling that the logical error rate scales exponentially with $L$, we expect the optimal logical error rate to be
\begin{align}
\log(1/\overline{p}_*)\simeq c'n_e^{1/2} \eta^{-1/2}+c'',\label{eq:ansatzn}
\end{align}
where $c'$ and $c''$ are positive constants. On the other hand, we can also consider the case in which $n_e$ scales linearly with $L$, i.e., $n_e = L/m$ for some integer $m$. In this case, the maximum length one can have is $L\propto m^{-1}\eta^{-1}$. Then the optimal logical error rate would be
\begin{align}
\log(1/\overline{p}_*) \simeq c'm^{-1}\eta^{-1} +c''.\label{eq:ansatzm}
\end{align}

\begin{figure*}
\centering
\begin{overpic}[width=0.3\textwidth]{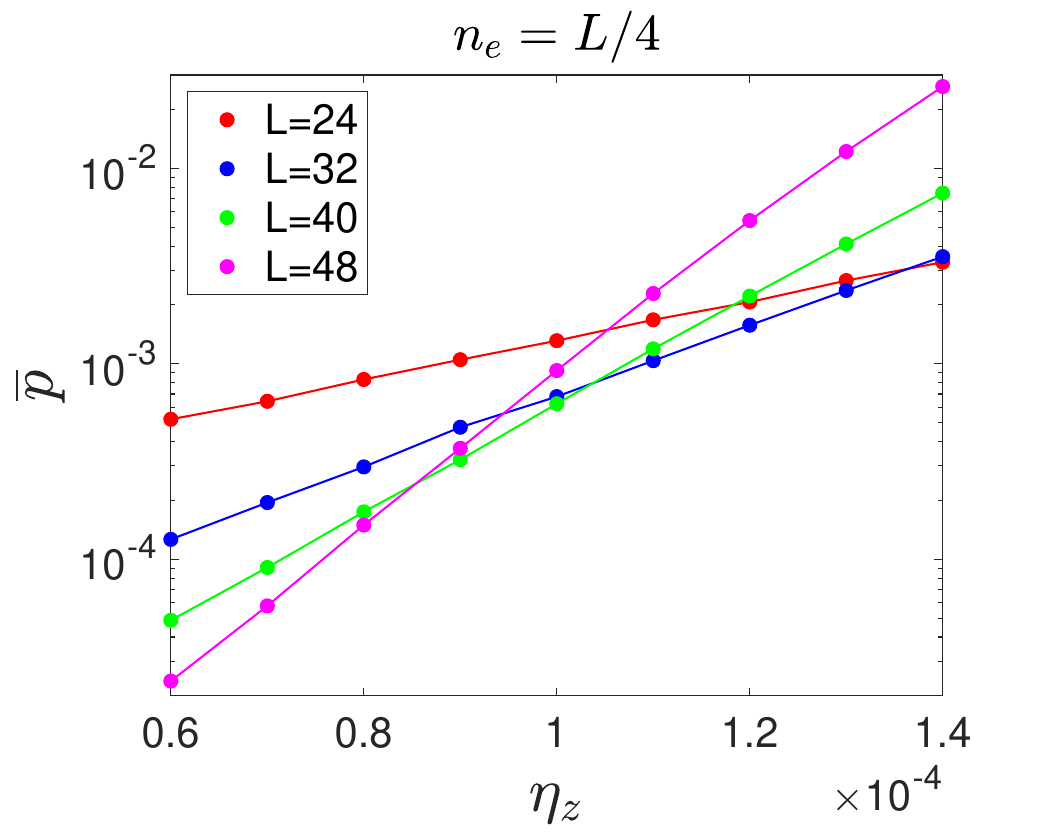}
\put(-5,75){(a)} 
\end{overpic}
\begin{overpic}[width=0.3\textwidth]{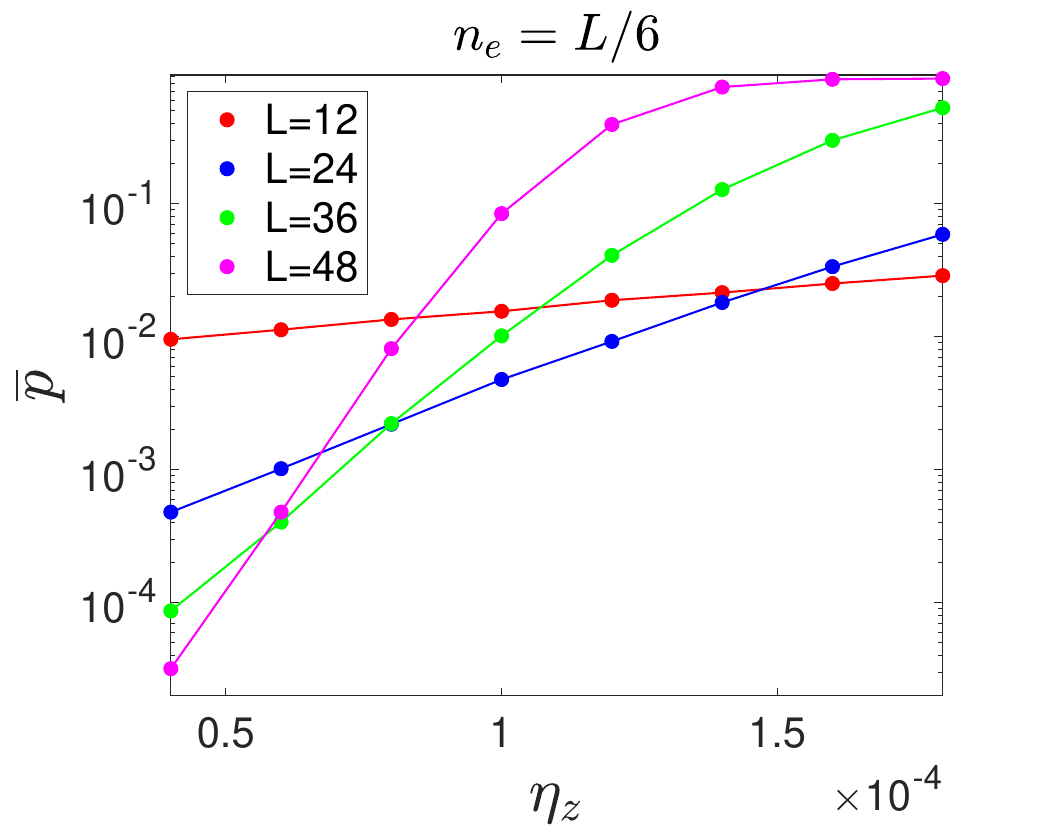}
\put(-5,75){(b)} 
\end{overpic}
\begin{overpic}[width=0.3\textwidth]{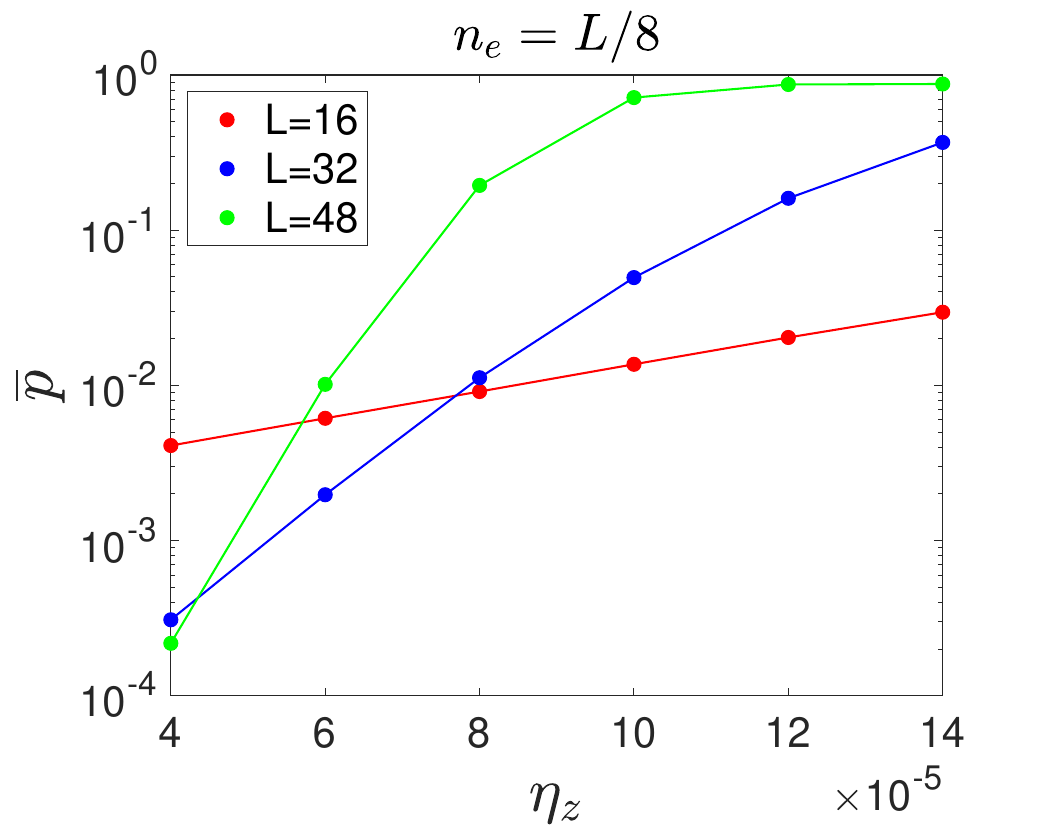}
\put(-5,75){(c)} 
\end{overpic}
\\
\begin{overpic}[width=0.3\textwidth]{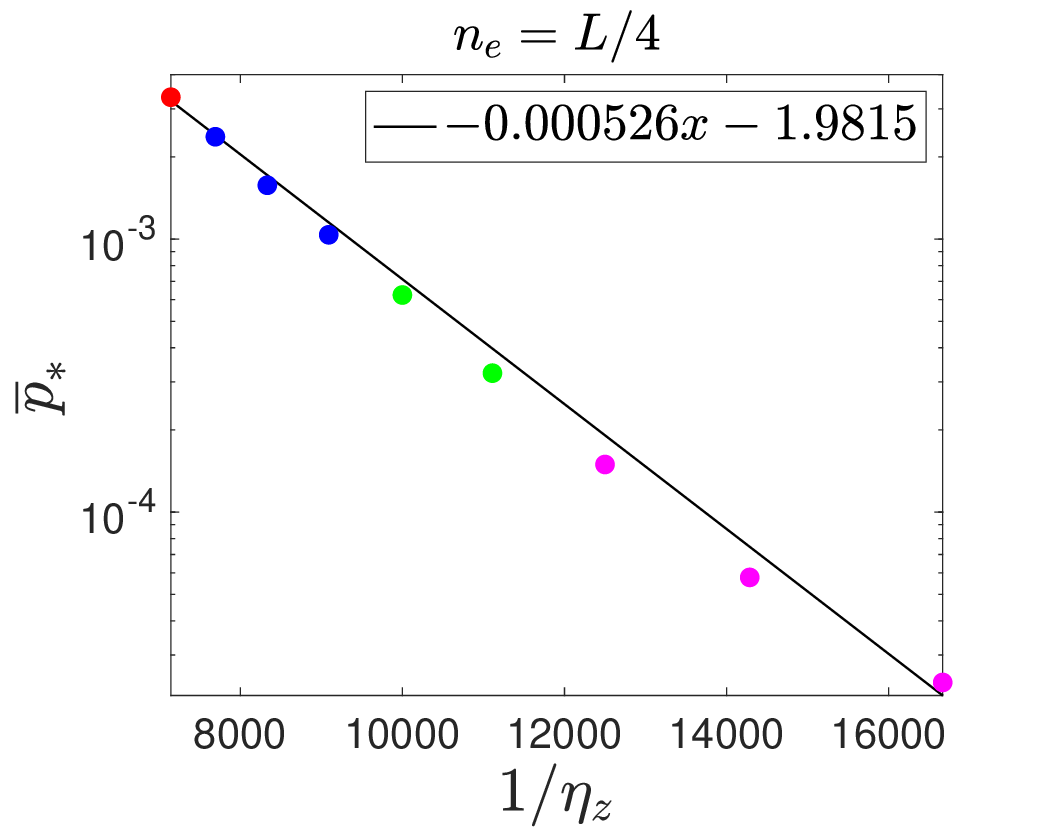}
\put(-5,75){(d)} 
\end{overpic}
\begin{overpic}[width=0.3\textwidth]{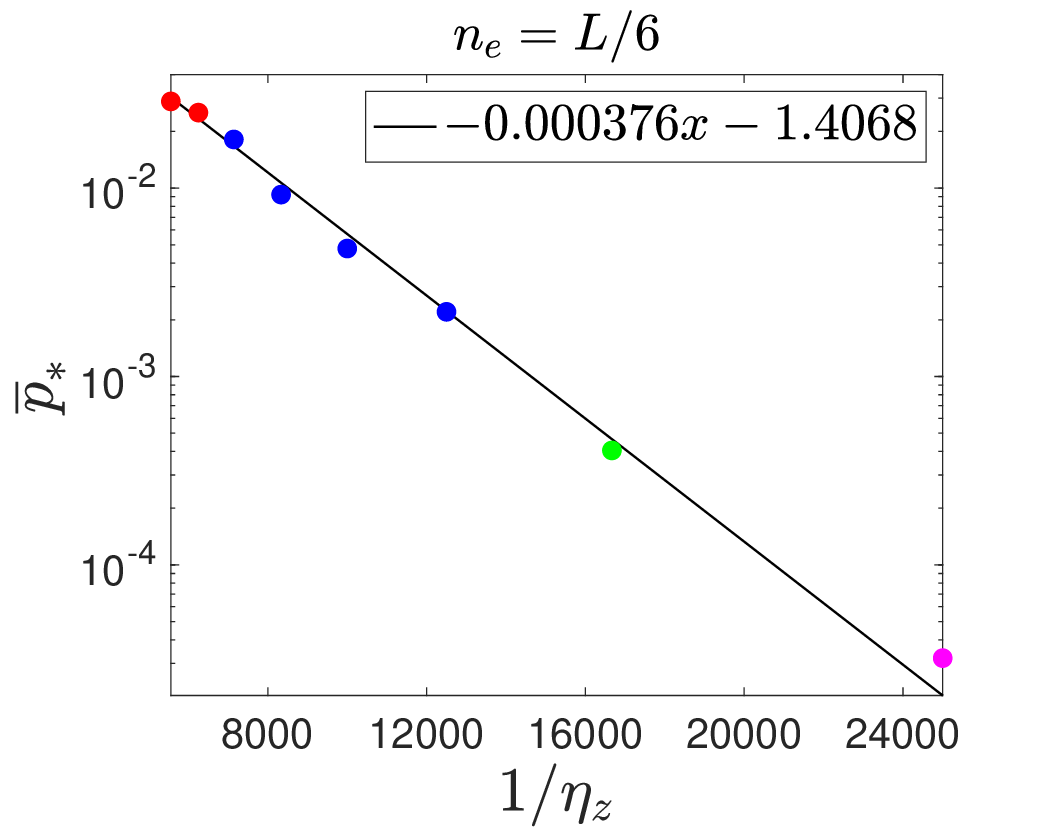}
\put(-5,75){(e)} 
\end{overpic}
\begin{overpic}[width=0.3\textwidth]{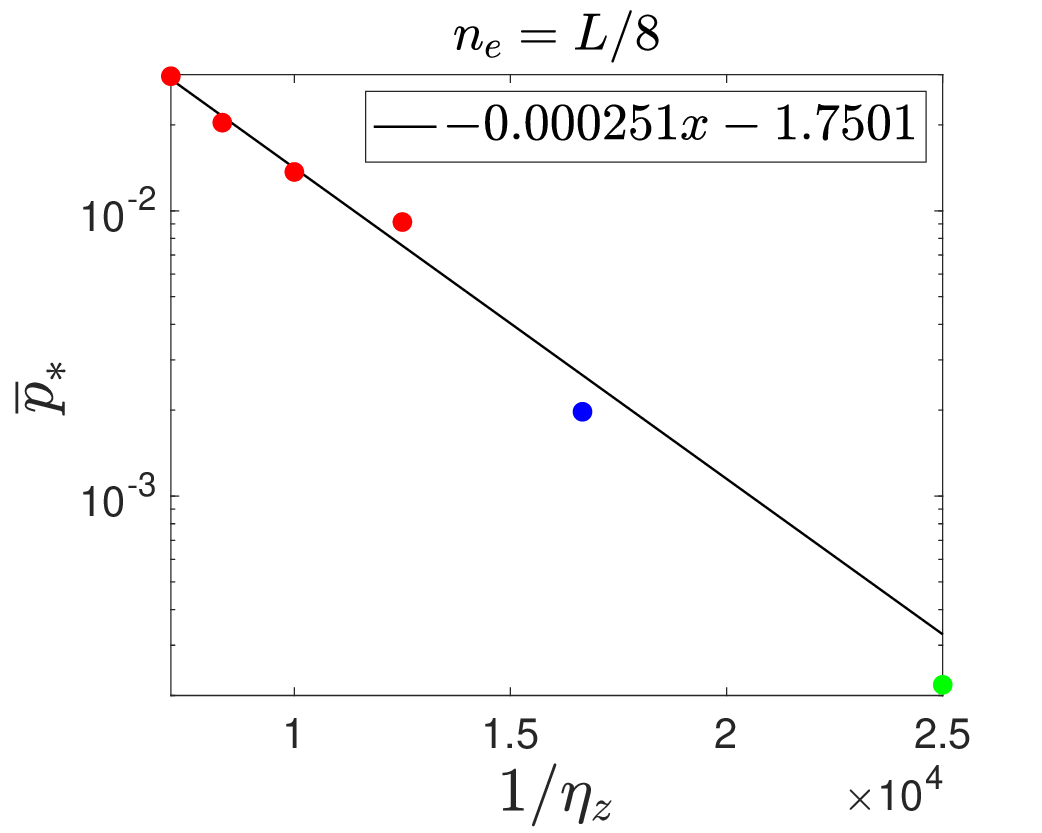}
\put(-5,75){(f)} 
\end{overpic}
\caption{Top panel: logical error rates versus dephasing error rates for (a) $L/4$ emitters (b) $L/6$ emitters (c) $L/8$ emitters, employing Protocol \Mref{alg3}. Bottom panel: $\overline{p}_*$ and its ansatz in Eq. \eqref{eq:ansatzm} versus $\eta_z^{-1}$ for (d) $L/4$ emitters (e) $L/6$ emitters (f) $L/8$ emitters, employing Protocol \Mref{alg3}. The color of the dots represents the value of $L$ corresponding to the minimum $\overline{p}$ for a given $\eta_z$, with the coloring scheme following the legend in the upper figures. The $y$-axis are on a logarithmic scale.}
\label{fig:9}
\end{figure*}

Now we discuss the results of our simulations. We will primarily focus on the effect of dephasing error, and only briefly mention about the loss error simulation at the very end of this Section. We first comment on the case of $n_e = O(1)$ [Eq.~\eqref{eq:ansatzn}]. We considered $n_e = 2, 4, 6$ using $10^5$, $10^6,$ and $10^7$ samples per data point, using the circuit-level depolarizing noise of $p=10^{-3}$. (Larger number of $n_e$ required larger sample size because the logical error rates obtained is lower.) All these data points were well-described by the values of $c'\approx 2.0 \times 10^{-2}$ and $c''\approx 1.4$ for \Mref{alg3} and $c' \approx 3.0\times 10^{-2}$ and $c''\approx 1.7$ for \Mref{alg4}, respectively; see Figure~\ref{fig:7} and~\ref{fig:8}.\footnote{More precisely, for \Mref{alg3}, we obtained $c'=0.0215$, $0.0204$, $0.0204$ for $n_e=2$, $4$, $6$, respectively. For \Mref{alg4}, we obtained $c'=0.0288$, $0.0302$, $0.0329$ for $n_e=2$, $4$, $6$, respectively.} We remark that the values of $c'$ and $c''$ for algorithms \Mref{alg3} and \Mref{alg4} are similar to those in \ref{alg1} and \ref{alg2}, respectively (corresponding to the $n_e=1$ case).   Therefore, we can conclude that the ansatz in Eq. \eqref{eq:ansatzn} adequately describes the scaling behavior of the logical error rate. Using this equation, we listed the requisite dephasing error rates to achieve targeted logical error rates of $\overline{p}_*=10^{-3}, 10^{-5}, 10^{-10}, 10^{-15}$ are summarized in Table \ref{tab:2} (top), (bottom).

\begin{table}[ht]
\footnotesize
\centering
\begin{tabular}{c| c| c| c}
\hline 
\hline 
\multirow{2}{*}{\backslashbox{$\overline{p}_*$}{\Mref{alg3}}}& \multirow{2}{*}{$2$ ($\eta_z$)} & \multirow{2}{*}{$4$ ($\eta_z$)} & \multirow{2}{*}{$6$ ($\eta_{z}$)} \\
& & & \\
\hline
$10^{-3}$ & $2.91 \times 10^{-5}$ & $5.66 \times 10^{-5}$ & $8.32 \times 10^{-5}$\\
\hline
$10^{-5}$ & $8.81 \times 10^{-6}$& $1.65 \times 10^{-5}$ &  $2.45 \times 10^{-5}$\\
\hline
$10^{-10}$ & $1.95 \times 10^{-6}$ & $3.57 \times 10^{-6}$ & $5.34 \times 10^{-6}$\\
\hline
$10^{-15}$ & $8.35 \times 10^{-7}$ & $1.51 \times 10^{-6}$& $2.27 \times 10^{-6}$\\
\hline
\hline
\end{tabular}

\centering

\vspace*{10pt}

\begin{tabular}{c| c| c| c}
    \hline 
    \hline 
\multirow{2}{*}{\backslashbox{$\overline{p}_*$}{\Mref{alg4}}}& \multirow{2}{*}{$2$ ($\eta_z$)} & \multirow{2}{*}{$4$ ($\eta_z$)} & \multirow{2}{*}{$6$ ($\eta_{z}$)} \\
& & & \\
\hline
$10^{-3}$ & $6.37 \times 10^{-5}$ & $1.43 \times 10^{-4}$ & $2.10 \times 10^{-4}$\\
\hline
$10^{-5}$ & $1.76 \times 10^{-5}$ & $3.91 \times 10^{-5}$ & $6.29 \times 10^{-5}$\\
\hline
$10^{-10}$ & $3.69 \times 10^{-6}$ & $8.13 \times 10^{-6}$ & $1.38 \times 10^{-5}$\\
\hline
$10^{-15}$ & $1.55 \times 10^{-6}$ & $3.41 \times 10^{-6}$ & $5.89 \times 10^{-6}$\\
\hline
\hline
\end{tabular}
\caption{The requisite dephasing error rates for (top) Protocol \Mref{alg3} employing $2,4,6$ emitters (bottom) Protocol \Mref{alg4} employing $2,4,6$ emitters to achieve targeted logical error rates of $\overline{p}_*=10^{-3}, 10^{-5}, 10^{-10}, 10^{-15}$ are listed. 
} 
\label{tab:2}
\end{table}

Under the assumption that the optimal choice of $L$ is proportional to $\sqrt{n_e/\eta}$, the requisite $L$ for the $L \times L \times L$ 3D cluster state to achieve $\overline{p}_* = 10^{-3}, 10^{-5}, 10^{-10}, 10^{-15}$ for dephasing error employing Protocol \Mref{alg3} with $n_e=2$ are approximately $25$, $50$, $100$, and $155$, respectively. Similarly, for $n_e = 4$, the requisite values of $L$ are approximately $25$, $45$, $100$, and $145$, respectively. For $n_e = 6$, the required values of $L$ are approximately $25$, $50$, $100$, and $160$, respectively.

For dephasing error employing Protocol \Mref{alg4} with $n_e=2$, the requisite $L$ are approximately $18$, $35$, $75$, and $115$, respectively. Similarly, for $n_e = 4$, the requisite values of $L$ are approximately $17$, $33$, $70$, and $110$, respectively. For $n_e = 6$, the required values of $L$ are approximately $18$, $32$, $70$, and $105$, respectively.

For the case of $n_e = L/m$, we chose $m= 4, 6, 8$ and employed Protocol \Mref{alg3} [Figure~\ref{fig:9}].\footnote{While we carried out the same simulation for Protocol \Mref{alg4} as well, even with $10^7$ samples we could not determine the minimum logical error rate $\overline{p}_*$. In particular, we could not obtain the constants in Eq.~\eqref{eq:ansatzm}. Nonetheless, we found the break-even point with $n=L/2$ to be greater than $2.9\times 10^{-3}$.} Examining the plots presented in Figure \ref{fig:9} (bottom), we observe a linear relationship between $\log \overline{p}_*$ and $\eta_{z}^{-1}$ for $n_e=L/4, L/6, L/8$. The values of $c'$ and $c''$ obtained were $c' \approx 2.1 \times 10^{-3}$ and $c'' \approx 1.7$ for all the cases.\footnote{More precisely, we obtained $c'=0.002104,$ $ 0.002256,$ $0.002008$ for $m=4,~ 6,~ 8$.} The requisite dephasing error rates to achieve targeted logical error rates of $\overline{p}_*=10^{-3}, 10^{-5}, 10^{-10}, 10^{-15}$ are summarized in Table \ref{tab:3}.

\begin{figure*}
\centering
\small
\begin{overpic}[width=0.3\textwidth]{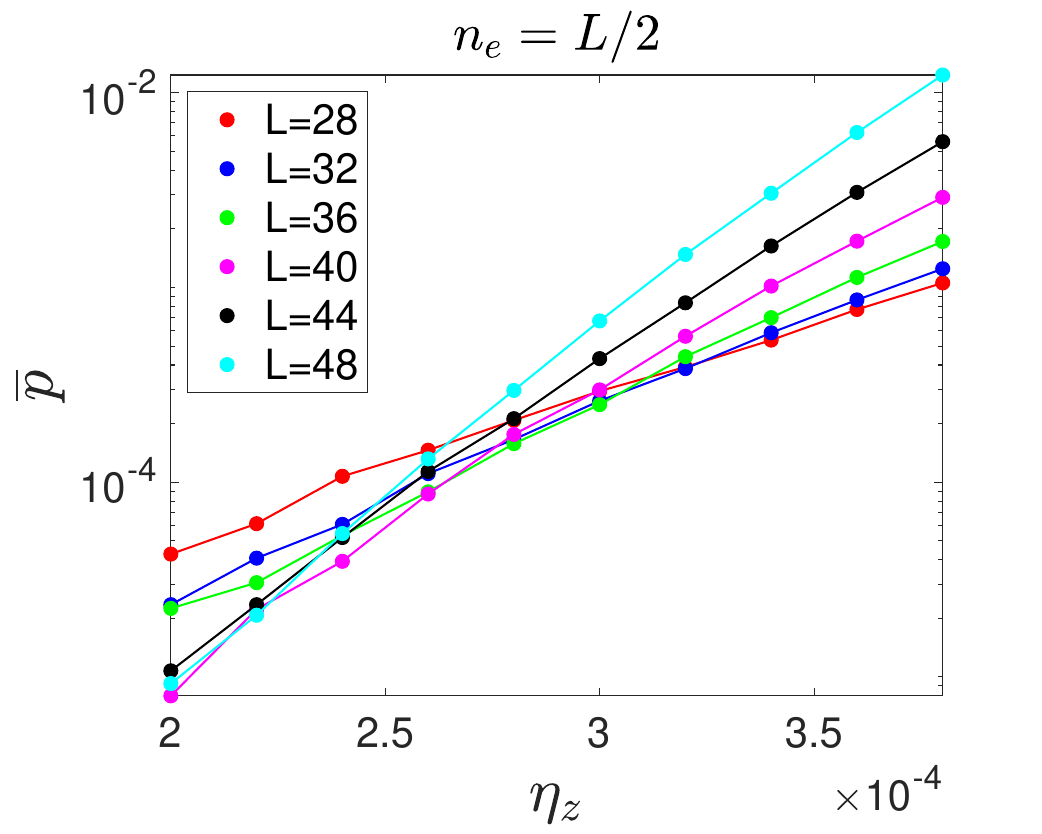}
\put(-5,75){(a)} 
\end{overpic}
\begin{overpic}[width=0.3\textwidth]{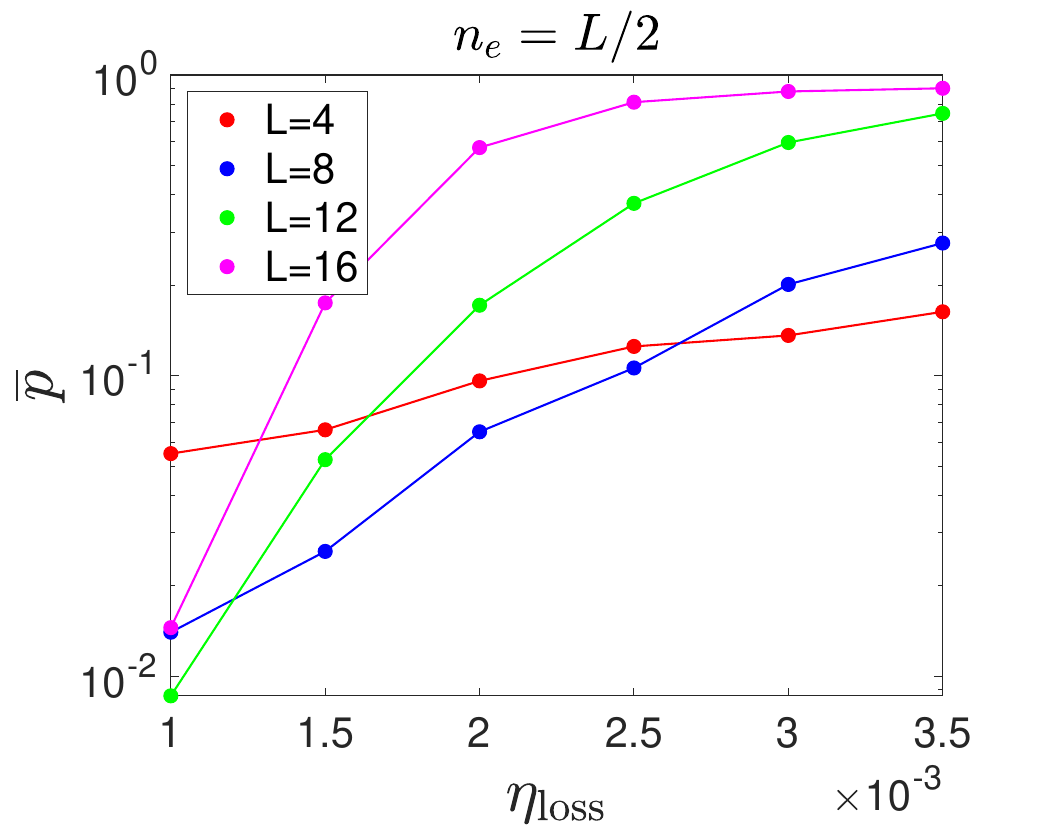}
\put(-5,75){(b)} 
\end{overpic}
\\
\begin{overpic}[width=0.3\textwidth]{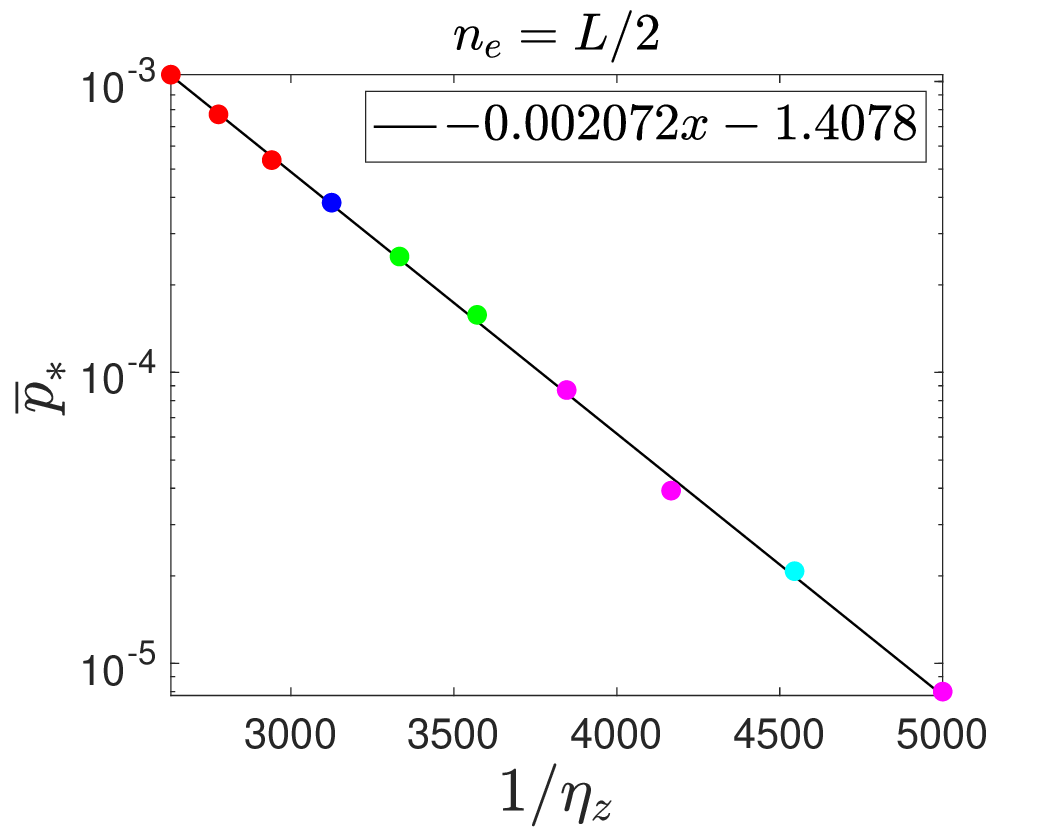}
\put(-5,75){(c)} 
\end{overpic}
\begin{overpic}[width=0.3\textwidth]{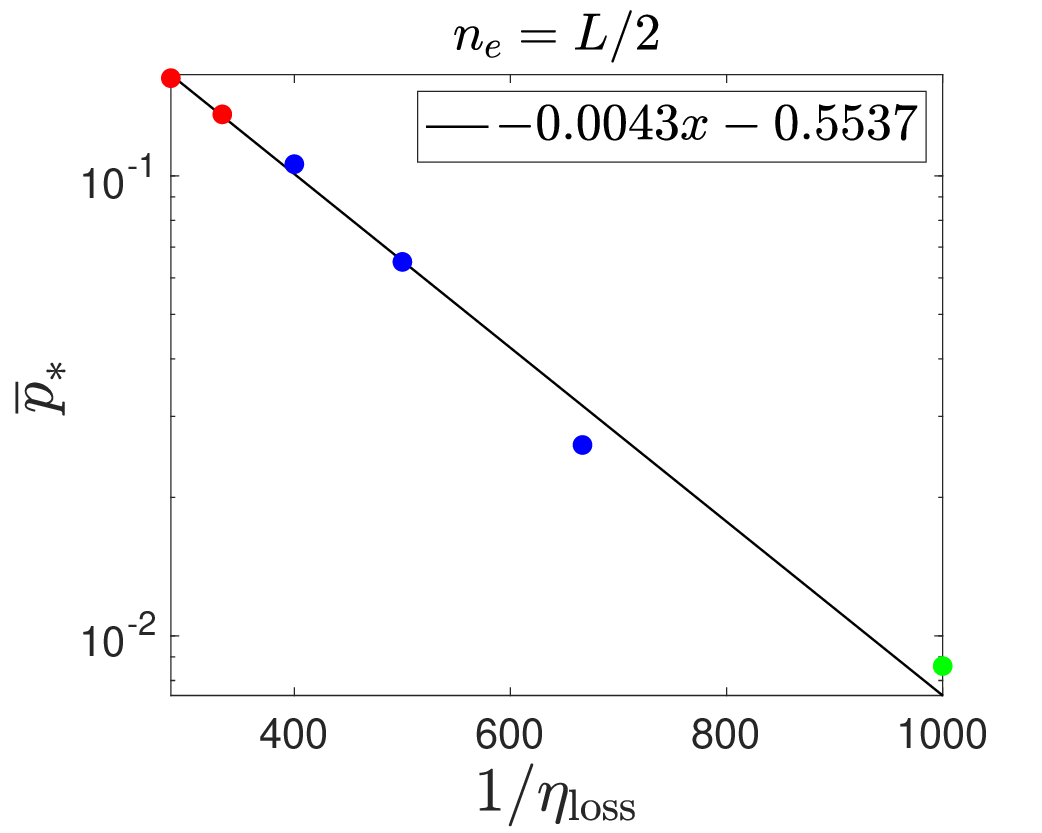}
\put(-5,75){(d)} 
\end{overpic}
\caption{Top panel: logical error rates versus (a) dephasing error rates for $L/2$ emitters and (b) loss error rates for $L/2$ emitters. Bottom panel: $\overline{p}_*$ and its ansatz in Eq. \eqref{eq:ansatzm} versus (c) $\eta_{z}^{-1}$ for $L/2$ emitters, and (d) $\eta_{\rm loss}^{-1}$ for $L/2$ emitters. The color of the dots represents the value of $L$ corresponding to the minimum $\overline{p}$ for a given $\eta_z$, with the coloring scheme following the legend in the upper figures. The $y$-axis are on a logarithmic scale.
}
\label{fig:10}
\end{figure*}

Under the assumption that the optimal choice of $L$ is proportional to $m^{-1}\eta^{-1}$, the requisite $L$ for the $L \times L \times L$ 3D cluster state to achieve $\overline{p}_* = 10^{-3}, 10^{-5}, 10^{-10}, 10^{-15}$ for dephasing error employing Protocol \Mref{alg3} with $n_e=L/4$ are approximately $35$, $70$, $150$, and $230$, respectively. Similarly, for $n_e = L/6$, the requisite values of $L$ are approximately $30$, $55$, $115$, and $180$, respectively. For $n_e = L/8$, the required values of $L$ are approximately $25$, $50$, $110$, and $170$, respectively.

\begin{table}[h!]
\centering
\footnotesize
\begin{tabular}{c| c| c| c}
    \hline 
    \hline 
\multirow{2}{*}{\backslashbox{$\overline{p}_*$}{\Mref{alg3}}}& \multirow{2}{*}{$L/4$ ($\eta_z$)} & \multirow{2}{*}{$L/6$ ($\eta_z$)} & \multirow{2}{*}{$L/8$ ($\eta_{z}$)} \\
& & & \\
\hline
$10^{-3}$ & $1.07 \times 10^{-4}$ & $6.83 \times 10^{-5}$ & $4.87\times 10^{-5}$\\
\hline
$10^{-5}$ &$5.52 \times 10^{-5}$ & $3.72 \times 10^{-5}$ & $2.57\times 10^{-5}$\\
\hline
$10^{-10}$ & $2.50 \times 10^{-5}$ &$1.74 \times 10^{-5}$ & $1.18\times 10^{-5}$\\
\hline
$10^{-15}$ & $1.62 \times 10^{-5}$ & $1.13 \times 10^{-5}$ & $7.65\times 10^{-6}$\\
\hline
\hline
\end{tabular}
\caption{The requisite dephasing error rates for Protocol \Mref{alg3} employing $L/4$, $L/6$, $L/8$ emitters to achieve targeted logical error rates of $\overline{p}_*=10^{-3}$, $10^{-5}$, $10^{-10}$, $10^{-15}$ are listed.
} 
\label{tab:3}
\end{table}

We remark that for $m=2$, the result of our simulation was inconsistent with Eq.~\eqref{eq:ansatzm}; see Figure \ref{fig:10} (a) and (c). More precisely, we obtained $c'\approx 4.1\times 10^{-3}$, which is different from the other values of $m$ (which were $c'\approx 2.1 \times 10^{-3}$). This discrepancy can be inferred from the relationship between the number of qubits associated with the gates between the emitters, given by $4L^3/(3m)$, and the number of other qubits, expressed as $3L^3/4 - 4L^3/(3m)$. (See Figure~\ref{fig:4} for a numerical validation of these values.) For $m=2$, the former exceeds the latter, whereas for $m=4, 6, 8$, it falls below.  Therefore, we anticipate that for $2 < m < 4$, the value of $c'$ will gradually increase as $m$ decreases.

Lastly, let us briefly mention the result of our loss error simulation. Similar to the loss error simulation with a single emitter, we have removed the circuit-level noise. To that end, we considered the phenomenological noise model in Eq.~\eqref{eq:etaloss}, employing the decoding algorithm in Ref.~\cite{stace10}; see Figure~\ref{fig:10} (b) and (d). Each data point is obtained from $10^4$ samples. The only result we present here are the case of $n_e =L/2$, which yields the lowest possible logical error rate out of all possible choice of $n_e$ we discuss; see Section~\ref{sec:optimal_delay_line}.

\subsection{Comparison: single- vs. multi-emitter}
\label{sec:optimal_delay_line}

In this Section, we discuss what kind of improvements the multi-emitter protocol [Section \ref{sec:3}] provides over the single-emitter protocol [Section \ref{sec:2}]. Recall that the optimal logical error rate improves as $n_e$ increases [Section~\ref{sec:5}]. Since the largest possible value of $n_e$ is $n_e=L/2$, we focus on comparing this case to the single-emitter protocol.

In Table~\ref{tab:4}, we listed the dephasing and loss error rates needed to achieve the target logical error rates of $\overline{p}_*=10^{-3}, 10^{-5}, 10^{-10}, 10^{-15}$. Comparing these results to the simulation results for the single-emitter protocol [Table~\ref{tab:1} (top)], we observe an approximately a ten-fold improvement (for the dephasing error) and a five-fold improvement (for the loss error) for achieving the break-even point (defined as $\overline{p}_*=10^{-3}$). Even greater improvements are achieved for lower target logical error rates.

\begin{table}[h!]

\centering
\footnotesize
\begin{tabular}{c| c| c}
\hline 
\hline 
\multirow{2}{*}{\backslashbox{$\overline{p}_*$}{$L/2$}}& \multirow{2}{*}{dephasing ($\eta_z$)} & \multirow{2}{*}{loss ($\eta_{\rm loss}$)}\\
& & \\
\hline
$10^{-3}$ & $3.77 \times 10^{-4}$ & $6.85 \times 10^{-4}$\\
\hline
$10^{-5}$ & $2.05 \times 10^{-4}$& $3.97 \times 10^{-4}$\\
\hline
$10^{-10}$ & $9.58 \times 10^{-5}$ & $1.94 \times 10^{-4}$\\
\hline
$10^{-15}$ & $6.25 \times 10^{-5}$ & $1.28 \times 10^{-4}$\\
\hline
\hline
\end{tabular}

\caption{The optimal requisite dephasing error rates for Protocol \Mref{alg3} and the optimal requisite loss error rates to achieve targeted logical error rates of $\overline{p}_*=10^{-3}, 10^{-5}, 10^{-10}, 10^{-15}$ are listed.
} 
\label{tab:4}
\end{table}

Under the assumption that the optimal choice of $L$ is proportional to $m^{-1} \eta^{-1}$, the requisite $L$ for the $L \times L \times L$ 3D cluster state to achieve $\overline{p}_* = 10^{-3}, 10^{-5}, 10^{-10}, 10^{-15}$ for dephasing error with $n_e=L/2$ are approximately $27$, $50$, $105$, and $165$, respectively. Similarly, the requisite $L$ for loss error with $n_e=L/2$ are approximately $25$, $45$, $90$, and $140$. Compared to the requisite $L$ needed to achieve certain logical error rates for a single emitter, a much larger $L$ can be used for $n_e = L/2$ emitters to achieve the same logical error rates with noisier delay lines.

\section{Discussion}
\label{sec:6}

In this paper, we propose protocols for constructing a cluster state using a linear array of emitters. Our key observation is that the protocol in Ref.~\cite{wan21} can be generalized to a protocol involving multiple emitters with a simple modification: an intermittent application of \textsf{CZ} gates between pairs of emitters. 

Such a modification reduces the amount of time each photon travels in the delay line, thereby improving the logical error rate overall. Although having a larger number of emitters may be more challenging than having only a single emitter, there are several reasons to prefer the multi-emitter protocol over the single-emitter protocol. The primary reason is the improved logical error rate. If one compares the delay line error rate needed to achieve the same logical error rate, the multi-emitter protocol [Table~\ref{tab:4}] outperforms the single-emitter protocol [Table~\ref{tab:1}] by a factor of at least $5\sim 10$, depending on the error model and the target logical error rate. Second, our protocol enjoys a high tolerance against two-qubit gates applied between the emitters. The thresholds for such gates were shown to be at least ten-fold larger than the threshold for the other gates [Figure~\ref{fig:LL}]. Lastly, the number of emitters is a tunable parameter in our scheme, and even a modest improvement in the number of emitters immediately yield improvements in the logical error rate [Section~\ref{sec:two_emitters_generalization}].

An interesting question is whether an error suppression can be demonstrated using our scheme in a realistic experiment. More specifically, we would define error suppression as an outcome in which the logical error rate after the error correction is lower than the physical error rate. Setting the physical error rate of $p=10^{-3}$, assuming the dominant source of error in the delay line is loss, we would obtain the break-even point ($\overline{p}_*=10^{-3}$) of $\eta_{\text{loss}}=6.85\times 10^{-4}$. This is lower than what the state-of-the-art delay line can achieve, e.g., $\eta_{\text{loss}} = 9.6\times 10^{-4}$~\cite{Tamura2018}. Therefore, our scheme, in the form presented in this paper, will not be able to achieve error suppression, even if we assume we used the state-of-the-art delay lines.

However, we think there are a few simple modifications in the protocol that can make the break-even point (for the delay line error) higher. Most importantly, we think simply changing our boundary condition to the open boundary condition will improve the result substantially. To see why, let us remark that our setup differs from that of Ref.~\cite{wan21}; the latter used an open boundary condition whereas we used a periodic boundary condition. The details about the loss error model is also different. While we obtained a similar threshold in spite of these differences, their sub-threshold behaviors are more markedly different; we found the loss error rates needed in our protocol to achieve specific target logical error rates were \emph{lower} than that of Ref.~\cite{wan21} by a factor of $3\sim 5$.

What would happen if we change the boundary condition to the open boundary condition? We conjecture that the scaling form of the optimal logical error rate [Eq.~\eqref{eq:ansatzn} and~\eqref{eq:ansatzm}] would be still valid even under such a condition, though with different constants. Let us briefly justify it. First, we remark that we are assuming that the circuit-level noise remains at $p=10^{-3}$, which is below the threshold (in the absence of other error sources). The additional error comes from the qubit loss, which is proportional to $L^2/n_e$ for every qubit, for both error models. While the details about this loss error model is different, this difference only causes a few additional gates for each data qubit. More precisely, in Ref.~\cite{wan21}, the loss can occur during the construction of the cluster state whereas in our case the loss occurs at the very end of the protocol. When qubit loss occurs during the construction of the cluster state, all subsequent operations acting on the lost data qubit become identity operators~\cite{wan21}. Therefore, in the entire procedure of gate operations, the loss error model where qubit loss occurs at the very end of the protocol requires only a few additional gate operations compared to the model where qubit loss occurs during the construction of the cluster state. While these are clearly different error models, they can be both described by some local error model. As such, we do not expect the logical error rate scaling form to be different.

The constants in Eq.~\eqref{eq:ansatzn} can be inferred from the Ref.~\cite{wan21}. The scaling relation in Ref.~\cite{wan21} would correspond to the $n_e=1$ case in Eq.~\eqref{eq:ansatzn}, yielding $c'=0.096$ and $c''=3.37$. Then, even with the use of two emitters ($n_e=2$), the requisite loss error rate to reach the break-even point becomes $\eta_{\rm loss}=1.47\times 10^{-3}$, which is strictly larger than the reported error rate of $\eta_{\rm loss}=9.6 \times 10^{-4}$~\cite{Tamura2018}. Therefore, provided that our assumption on the error scaling is correct, we are led to the conclusion that error suppression can be demonstrated with just two emitters and four delay lines. This is more demanding than what was originally envisioned in Ref.~\cite{wan21}, but only barely.

Of course, the picture we described so far is too simplistic. For one thing, there can be an additional insertion error occurring as the photon moves in and out of the delay line. Moreover, there are other schemes, such as the one using concatenation, which were shown to improve the noise tolerance~\cite{Li2023concatenation}. In order to more accurately assess the experimental prospect of our multi-emitter protocol, it is important to accurately model these sources of error and employ the error correction schemes that are adept at correcting such errors.

Lastly, we remark that a more refined comparison between the periodic boundary condition (used by us) and the open boundary condition (used in Ref.~\cite{wan21}) would be interesting. On one hand, the open boundary conditions appears to better suppress errors. On the other hand, periodic boundary condition can encode twice as more qubits; this is because latter can be viewed as a foliation of the 2D toric code~\cite{stace16}.) Which would be better suited for building a fault-tolerant quantum computer? We leave this question for future work.


\textbf{Note}: The source codes to reproduce the numerical results are available on Zenodo~\cite{zenkim}.

\acknowledgments 
We thank Hassan Shapourian and Alireza Shabani for useful discussions and Yun-Tak Oh for helping with simulations. J. K. was supported by the education and training program of the Quantum Information Research Support Center, funded through the National research foundation of Korea (NRF) by the Ministry of science and ICT (MSIT) of the Korean government(No.2021M3H3A103657313). 
J.H.H. was supported by the National Research Foundation of Korea (NRF) grant funded by the Korea government (MSIT) (No. 2023R1A2C1002644). I. K. was supported by the Cisco Research Gift Program.




  
\bibliographystyle{plainnat}
\bibliography{SC}

\end{document}